%% file: main.tex
\def\runningtitle{Source exclusion in regime-gated forecasting}
\input{preamble.tex}
\hypersetup{pdftitle={Source Exclusion in Regime-Gated Forecasting: A Cross-Market Audit of Equity-Factor Predictability},pdfauthor={Chorok Lee}}
\title{Source Exclusion in Regime-Gated Forecasting:\\A Cross-Market Audit of Equity-Factor Predictability}
\author{Chorok Lee\\Korea Advanced Institute of Science and Technology}
\date{}
\begin{document}
\maketitle
\begin{abstract}
A source variable can enter a forecast through both explicit lag coefficients and an estimated regime gate. Removing only its lag block therefore evaluates a different procedure from excluding its information throughout fitting, filtering and tuning. We audit this established distinction in equity-factor forecasting. An initial current-vintage reconstruction of the reported HML-to-SMB relation finds negative incremental MSE gains for pooled, Gaussian-HMM and Student-$t$-HMM forecasts in 2010--2024. A separately frozen expansion evaluates all 30 factor directions in four regional panels, with 2025--August 2026 withheld from earlier repository forecast evaluation. Shared-gate ablation and complete exclusion give opposite capped-loss point-estimate signs in 42 of 240 unweighted temporal comparisons. None of the 2,400 temporal endpoints, including predictable regime weights, has a directional crossing under an established simultaneous confidence-sequence construction. All 776 intervals within a declared score-resolution margin are already explained by forecast proximity. An initial 400 scenario-replications with known conditional means separate removal contrasts from procedure costs; a further 400 paired noise-level replications show fitted HMM losses despite positive oracle information. A matched Gaussian-versus-Bernoulli channel intervention on 400 further paired scenario-replications improves forecasts at one corruption level under known-AR residualization, without identifying the earlier pipeline failure or a financial mechanism. Application-budget and learned-pipeline diagnostics expose poor resolution of known positive targets. Independent implementations replay forecasts from archived HMM fits and recompute all 4,800 primary financial interval summaries. The evidence demonstrates sensitivity to the removal procedure and limited inferential resolution, rather than broad predictive success or population noncausality. Correlated markets, retrospective data vintages and unverified outside exposure limit confirmation claims.
\end{abstract}

\section{Introduction}
Factor comovement changes during market stress, but contemporaneous correlation cannot identify whether lagged information improves a forecast. Granger predictability asks whether the past of one series contributes to predicting another, conditional on a specified information set \cite{granger}. It is not an intervention-based causal claim.

The earlier preprint \cite{original} investigates whether the HML--SMB relationship varies across regimes estimated with a Student-$t$ hidden Markov model. Its reported results are potentially interesting, but a defensible contribution requires separating three questions: whether a relation exists in a retrospectively selected sample; whether the relation differs statistically across regimes; and whether it improves forecasts when both regime identification and parameter estimation use only information available at the forecast origin.

This revision replaces claims of established lead time with a chronological reassessment and a complete regional task family. It asks whether the removal operation changes measured incremental predictive value, and whether that value can be resolved under a stated inference target. The original US pair remains negative in evaluated point estimates, while the broader family includes positive, negative and zero estimates. The contribution is the measured cross-market comparison, controlled mechanism diagnostics and independently checked implementation. Heavy-tailed HMMs are established tools \cite{bulla}; neither their use nor the sequential confidence construction below is claimed as a new method.

\paragraph{Position relative to forecasting literature.}
The separation of observation-equation and regime-inference restrictions follows established switching-model work \cite{droumaguet}. Forecast comparisons also distinguish parameter-estimation cost from population predictive content: Clark and West \cite{clarkwest} study nested-model MSPE comparisons, while Giacomini and White \cite{giacomini} develop conditional predictive-ability evaluation of fitted procedures. Their tests are not automatically valid for our expanding-window, tuned, estimated-gate pipeline. Our contribution is the documented chronological reassessment and its controls, rather than a new noncausality framework or a new test.

The expansion asks a more general empirical question: how much does measured source value change when removal covers the entire fitted pipeline rather than only its explicit lag block? Removal-based explanations already distinguish the removal operation from the behavior being explained \cite{covert}. Neural Granger methods use architecture-specific structured penalties to enforce input nondependence \cite{tank}; our concern is not a universal flaw in nonlinear Granger methods. The audit concerns a particular two-stage regime-gated design.

\begin{table}[ht]
\centering\small
\begin{tabular}{p{0.25\linewidth}p{0.34\linewidth}p{0.32\linewidth}}\toprule
Component & Established foundation & Role of this revision\\\midrule
Source restrictions & Switching-model observation and regime-inference restrictions \cite{droumaguet} & Explicitly audited fitted gate, filtering, lag and tuning paths\\
Removal interpretation & Choice of removal operation and explanatory target \cite{covert} & Matched shared-gate and complete-exclusion comparisons\\
Sequential comparison & Conditional-average score CSs \cite{choe2024,howard2021}; sequential model sets \cite{arnold2026} & Bounded-loss application with predictable paired ranges and a complete family\\
Empirical evidence & Earlier selected HML--SMB association \cite{original} & New chronological reconstruction, complete regional task family and separately recorded temporal extension\\\bottomrule
\end{tabular}
\caption{Contribution boundary. The revision contributes measured evidence and an auditable implementation; the underlying restriction and confidence-sequence principles are established.}\label{tab:position}
\end{table}

\section{Data and information sets}
The historical study reports 8,817 daily observations from 1990--2024 for MKT-RF, SMB, HML, RMW, CMA, and MOM. These are the Fama--French five factors plus momentum, joined from documented series in the French Data Library \cite{french}. Replication must retain the exact vintage, trading calendar, percentage-to-decimal conversion, missing-value handling, and SMB definition. Counts should be generated directly from the final analysis index.

For the new reconstruction, we downloaded the daily five-factor and momentum archives on 26 September 2026. Both identify the August 2026 CRSP database. An exact-date join from 2 January 1990 to 31 December 2024 gives 8,817 rows with no unmatched momentum dates in the selected five-factor calendar. We use the five-factor SMB series, omit RF as a predictor, rename the market factor MKT-RF, and divide percentage returns by 100. No dates are imputed. Raw archives and a hash manifest are retained. Matching the earlier observation count does not verify equality to its historical data vintage. This first reconstruction excludes values after 2024. A separately frozen expansion below evaluates three additional regional panels and extends the evaluation through August 2026.

The reconstructed SMB is the five-factor size spread, averaging size spreads from book-to-market, profitability and investment sorts; it differs from three-factor SMB. HML is the value-minus-growth spread. Their underlying stock portfolios overlap. Removing HML's numerical history is therefore an information-input restriction, not removal of every value-related economic channel or an intervention on independent portfolios. The original factor definition and vintage cannot be certified solely from the matching observation count.

Let $\mathbf r_t\in\R^6$ denote returns on trading day $t$, and let $\mathcal F_t$ contain research observations indexed through that day. Real-time use must additionally satisfy the publication-availability restrictions below. When predicting $r_{j,t}$, features must be measurable with respect to $\mathcal F_{t-1}$. When predicting $r_{j,t+h}$ at the end of $t$, the permitted information is $\mathcal F_t$. These time conventions apply to normalization, regime inference, hyperparameters, and any policy decision.

\section{Regime estimation without future observations}
Let $z_t\in\{1,\ldots,K\}$ have transition probabilities $A_{jk}$. The implemented gates use diagonal-scale Gaussian or Student-$t$ emissions; the latter fixes $\nu=5$. Appendix~\ref{app:emissions} gives the emission density and fitting details.

For frozen parameters $\widehat\theta$, the predictive and filtered probabilities are
\begin{align}
 w_{t|t-1,k}&=\sum_j A_{jk}w_{t-1|t-1,j},\\
 w_{t|t,k}&=\frac{p(\mathbf x_t\mid z_t=k;\widehat\theta)w_{t|t-1,k}}
 {\sum_m p(\mathbf x_t\mid z_t=m;\widehat\theta)w_{t|t-1,m}}.
\end{align}
The smoothed probability $\Prb(z_t=k\mid\mathbf x_{1:T})$ uses future observations whenever $t<T$. These quantities are not interchangeable. Even a forward filter leaks future information if its parameters or scaling were estimated on the full evaluation period.

State ranks use increasing training covariance trace; they are not verified economic labels. Parameters and scaling are fitted strictly before each evaluation block, and only revealed observations update the filter. Current-vintage research observations are not certified as available at their historical publication times (Appendix~\ref{app:availability}).

\section{Source-removal paths and predictive estimands}
\subsection{The fitted gated predictor}
The implemented gated linear predictor has the form:
\begin{equation}\label{eq:gated}
 r_{j,t}=\sum_{k=1}^K w_{t|t-1,k}
 \left[a_{j,k}+\sum_{\ell=1}^p\mathbf B_{j,k,\ell}\mathbf r_{t-\ell}\right]+\varepsilon_{j,t}.
\end{equation}
This is an explicit predictive regression with estimated regime weights, not automatically the exact conditional mean of every HMM. For a coefficient-ablation comparison, restricted and unrestricted models use identical regime weights and training windows; the restricted model sets the selected cross-factor lag coefficients to zero. This restriction alone is not an information-exclusion test when the gate uses the source factor. Including other factors supplies a stronger conditional information set than a bivariate regression.

\subsection{Two routes by which a source factor can predict the target}
The source factor may enter both the regression and the estimated regime probabilities. This distinction is established in Markov-switching VAR research: Droumaguet, Warne, and Wo\'zniak \cite{droumaguet} analyze restrictions on both the observation equation and regime inference. Our fitted gated regression is a forecasting approximation rather than their full likelihood model, but must respect the same information-set distinction.

Let $\mathcal G_{t-1}^{(-i)}$ contain the histories of all retained factors except source $i$, and let $\mathcal G_{t-1}$ additionally contain its history. Mean noncausality concerns
\begin{equation}
 \E[r_{j,t}\mid\mathcal G_{t-1}]
 =\E[r_{j,t}\mid\mathcal G_{t-1}^{(-i)}]\quad\text{almost surely}.
\end{equation}
An estimated gate depending on source history need not be measurable with respect to $\mathcal G_{t-1}^{(-i)}$. Zeroing explicit source-lag coefficients therefore need not construct a forecast from the smaller information set.

\begin{example}[Information retained by a shared gate]
Let $X_t$ be independent across time, with a continuous distribution symmetric about zero, and define $w_t=(1+\exp(-X_{t-1}))^{-1}$. Let $Y_t=w_t+\epsilon_t$, where $\epsilon_t$ is independent and identically distributed with zero mean and finite variance, independently of the entire $X$ sequence. The gated predictor $w_t a_1+(1-w_t)a_2$ with $a_1=1,a_2=0$ predicts the conditional mean without any explicit source-lag coefficient. Nevertheless, $X_{t-1}$ predicts $Y_t$: the source-excluded conditional mean based on past $Y$ is $1/2$, whereas the source-informed mean is $w_t$. Under squared loss, the population risk reduction is $\operatorname{Var}(w_t)>0$.
\end{example}
The last assertion follows by expanding $\E[(Y_t-1/2)^2]-\E[(Y_t-w_t)^2]$; independence makes the cross term zero. This elementary counterexample illustrates why coefficient ablation can miss a predictive pathway. It is not a claim of a new noncausality theorem.

We consequently distinguish three comparisons. A \emph{shared-gate ablation} measures additional explicit source-lag value conditional on that gate. An \emph{information-exclusion comparison} removes the source from every input pathway, including gate fitting, filtering, preprocessing, and tuning of the restricted procedure. A \emph{gate-value comparison} compares gated and pooled forecasts using the same raw information. These are different endpoints; a gain in one cannot stand in for a gain in the others. In the HMM information-exclusion experiment, the restricted gate is fitted and filtered using retained variables only; different state labels need not be aligned to compare their forecasts, although coefficient comparisons require additional identification work.

\begin{table}[ht]
\centering\small
\begin{tabular}{lccc}\toprule
Source access & $U$ & shared-$R$ & $R$\\\midrule
HMM fit and filter updates & Yes & Yes & No\\
Explicit source lag columns & Yes & No & No\\
Gate-derived design/scaling & Source-informed & Source-informed & Source-excluded\\
Validation forecast procedure & Full & Shared-gate & Source-excluded\\\bottomrule
\end{tabular}
\caption{The target differs from the removed source. Every path uses the same outcome history and evaluation dates. A retained source-informed gate makes shared-$R$ inadmissible as a complete information-exclusion forecast.}\label{tab:paths}
\end{table}
The difference between the two realized gains is exactly $L(R)-L(\text{shared-}R)$, because their unrestricted forecasts coincide. This is a comparison of restricted forecasting procedures. Independently re-estimating a gate changes its representation, estimation error and regularization; the difference is not an additive causal attribution to a latent information channel. The mechanism experiment below uses known conditional means to separate this distinction from estimation costs.

\subsection{Information value and finite-procedure performance}
The following standard conditional-expectation identity fixes the interpretation of forecast losses; it is not a new theorem.
\begin{proposition}[Oracle and fitted predictive gains]
Let $Y\in L^2$ and $\mathcal G\subseteq\mathcal F$. Write $m_G=\E[Y\mid\mathcal G]$ and $m_F=\E[Y\mid\mathcal F]$. Then
\begin{equation}
 \E[(Y-m_G)^2]-\E[(Y-m_F)^2]=\E[(m_F-m_G)^2]\geq0.
\end{equation}
For square-integrable forecasts $f_G$ and $f_F$ measurable with respect to their corresponding information sets, the actual risk difference is
\begin{align}
 \E[(Y-f_G)^2]-\E[(Y-f_F)^2]
 &=\E[(m_F-m_G)^2]\\
 &\quad+\E[(m_G-f_G)^2]-\E[(m_F-f_F)^2].\nonumber
\end{align}
\end{proposition}
\begin{proof}
Expand $Y-m_G=(Y-m_F)+(m_F-m_G)$. The cross term has expectation zero because $\E[Y-m_F\mid\mathcal F]=0$. Apply the same orthogonality separately to $Y-f_G=(Y-m_G)+(m_G-f_G)$ and $Y-f_F=(Y-m_F)+(m_F-f_F)$ to obtain the second identity.
\end{proof}
The information sets in this identity include each procedure's fitting and hyperparameter-selection history and algorithmic randomization. The restricted history must itself exclude the source. No training--target independence is needed for the projection identity. The identity also holds conditional on a common sigma-field $\mathcal C\subseteq\mathcal G$; conditioning instead on unrestricted source-containing training data changes the exclusion estimand. Application to a chronological sample average requires a separately justified evaluation target and dependence assumptions.

A fitted augmented model can consequently have a negative gain even when the source contains population predictive information: its additional approximation or estimation error can outweigh the oracle gain. Conversely, a positive realized sample average does not by itself establish positive population gain. We report finite-procedure losses rather than equating their sign with acceptance or rejection of Granger noncausality.

\section{Chronological fitted-procedure comparison}
The experiments use initial training, chronological validation for penalty selection, and later evaluation with scheduled expanding refits. At each cutoff, HMM parameters, feature scaling and regression coefficients use strictly preceding rows. Historical features within that fit sample use its fitted HMM: they are in-sample generated features, not historical online forecasts. Validation and evaluation gates use forward predictive probabilities with preceding parameter fits. Already revealed observations can update the filter within a block.

The pooled and gated predictors use ridge regression, with training-only column scaling and a free ordinary intercept. Each path selects its penalty by validation MSE and freezes that choice during evaluation; scaling and fitted coefficients are recomputed at refits. The expanded experiment uses symmetric all-state blocks and additionally allows an expanding historical-mean candidate. Zero return provides a further low-complexity benchmark. Appendix~\ref{app:protocolnotes} preserves the earlier reconstruction's additional baselines, coding choices and evaluation guidance; the expanded schedule is specified below.

For matched forecasts define $d_t=L(r_{j,t},\widehat r^R_{j,t|t-1})-L(r_{j,t},\widehat r^U_{j,t|t-1})$. Positive average $d_t$ favors source augmentation. Comparisons share target dates, outcomes and tuning access. The source-excluded gate, regressors, scaling and tuning omit the designated source throughout. One-day mean prediction is the executed task; lag order does not establish a forecast or warning horizon.

\section{Sequential inference for bounded forecasting losses}
Squared-loss comparisons require additional tail assumptions for finite-sample
inference. We therefore implement a
separate bounded-loss endpoint using established confidence-sequence methods
\cite{howard2021,choe2024}. This is an application of existing sequential
forecast-comparison theory, not a new concentration inequality or a
distribution-free test of Granger noncausality.

\subsection{Conditional target and a predictable pairing bound}
Let $(\mathcal F_t)$ describe the sequentially revealed research information.
The forecasts $f_{R,t},f_{U,t}$ and any evaluation weight $w_t\in[0,1]$ must be
$\mathcal F_{t-1}$-measurable. Fix a positive squared-return cap $c$ before the
evaluation period, allowing a choice made from the initial training sample.
Define
\begin{align}
 \ell_c(y,f)&=\min\{(y-f)^2/c,1\},\\
 d_t&=\ell_c(Y_t,f_{R,t})-\ell_c(Y_t,f_{U,t}),\qquad
 \mu_t=\E[d_t\mid\mathcal F_{t-1}],\\
 W_n&=\sum_{t=1}^n w_t,\qquad
 \Delta_n=\frac{\sum_{t=1}^n w_t\mu_t}{W_n},\qquad
 \widehat\Delta_n=\frac{\sum_{t=1}^n w_td_t}{W_n}.
\end{align}
The two averages are undefined when $W_n=0$. Positive $\Delta_n$ means that
the source-augmented procedure has smaller average conditional \emph{capped}
loss along the evaluated trajectory. The target may vary over time and be
random; it is not a stationary population MSE difference. Multiplication by
$c$ restores capped squared-return units, not uncapped MSE.

The map $f\mapsto\ell_c(y,f)$ is $2/\sqrt c$-Lipschitz, since its piecewise
derivative has magnitude at most $2/\sqrt c$. Therefore the known-before-outcome
forecast separation supplies the bound
\begin{equation}
 |d_t|\le b_t=\min\{1,2|f_{R,t}-f_{U,t}|/\sqrt c\},\qquad
 V_n=\sum_{t=1}^n(w_tb_t)^2.
\end{equation}
This pairing bound can be substantially smaller than one when the procedures
produce similar forecasts. It is an algebraic bound, not an estimated
conditional variance.

\begin{proposition}[Application of a two-sided normal-mixture boundary]
Suppose the above conditions hold for an outcome-independent family of $M$
comparisons. Fix $\alpha\in(0,1)$ and $\rho>0$ before evaluation, and set
$\alpha_j=\alpha/M$. For each comparison define
\begin{equation}
 u_j(v)=\sqrt{(v+\rho)\log\!\left\{\frac{v+\rho}{\rho\alpha_j^2}\right\}},
 \qquad
 C_{j,n}=\left[\widehat\Delta_{j,n}-\frac{u_j(V_{j,n})}{W_{j,n}},
 \widehat\Delta_{j,n}+\frac{u_j(V_{j,n})}{W_{j,n}}\right].
\end{equation}
With probability at least $1-\alpha$, $\Delta_{j,n}\in C_{j,n}$ simultaneously
for all comparisons and all times having positive weight mass. No temporal
independence, stationarity, or moment assumption on the uncapped returns is
required. The frozen application uses $M=2400$; this guarantee applies only
if that family is fixed independently of the evaluated outcomes or covers
all outcome-selected choices relevant to the reported assertion.
\end{proposition}
Proof and calibration details appear in Appendix~\ref{app:sequential}.
The formula is already two-sided. The implementation intersects endpoints
with the pathwise support
$[-\sum_{t\le n}w_tb_t/W_n,\sum_{t\le n}w_tb_t/W_n]$; consequently identical
forecasts yield the exact interval $[0,0]$. Its stored radius is the
\emph{untruncated} normal-mixture radius. Regime weights are admissible when
computed from past-only fits and predictive probabilities. Contrasts between
two weighted averages require simultaneous component bounds or another
justified construction: final-sample denominators cannot silently be used as
predictable per-observation weights.

\subsection{Calibration and resolution limits}
The initial 12-scenario calibration uses 2,000 independent replications per scenario and known conditional risks. A separately frozen extension matches the application's $M=2400$ allocation and US temporal horizon $n=416$. It retains synthetic two-point outcomes, unit weights and correlated comparisons, rather than reproducing financial forecasts or effect magnitudes. Neither calibration has a family coverage failure at any of its evaluated prefixes; this finite-horizon check does not empirically prove the all-time guarantee. In the matched extension, weak-alternative positive crossing is zero throughout; strong-alternative detection is 67.80\% under independent observations, 46.40\% under adaptive variance, and zero under persistent and abrupt-change environments. These results show conservatism and substantial power limitations, not effective detection across settings. Appendix~\ref{app:matched-calibration} reports all settings, widths and Monte Carlo intervals.

A further diagnostic applies the same construction to the 400 existing mechanism ledgers below, retaining learned forecasts, validation choices, training-random caps and predictable weights. Analytic Gaussian-mixture capped risks supply the exact conditional targets. This is reuse of previously inspected simulations, not fresh confirmation. For 33 actually evaluated endpoints, both the application-sized allocation $M=2400$ and natural allocation $M=33$ have zero observed family failures in 100 seeds per scenario. Yet neither resolves a positive unweighted exact-oracle, Gaussian or Student-$t$ information-exclusion target in any seed of the three signal scenarios, although all those terminal targets are positive. Reducing the multiplicity budget alone does not resolve this failure. Appendix~\ref{app:learned-inference} distinguishes conditional targets, coverage and descriptive detection.

\subsection{Variance-adaptive diagnostic}
The preceding resolution failure motivates a separately recorded application of established empirical-Bernstein machinery \cite{howard2021,choe2024}. The construction uses the observed cumulative squared weighted differences, with predictable stopping at fixed range violations and a countable, preallocated tuning family. Appendix~\ref{app:variance-adaptive} gives its formula and proof, including both-tail and range-selection corrections. It is an adaptation, not a new confidence-sequence principle. The original normal-mixture financial results remain the primary analysis. Choosing a second method after inspecting those results does not supply new confirmatory coverage for reused financial outcomes, nor license selecting the narrower of the two intervals without an additional method correction.

On the same 400 reused mechanism ledgers, the frozen variance-adaptive method yields positive crossings in only three mixed-scenario datasets at $M=2400$, all from pooled complete-exclusion comparisons; neither fitted HMM crosses. At $M=33$, some endpoint crosses in 50/100 lag-only and 86/100 mixed datasets, but none in gate-only. Thus reduced multiplicity and variance adaptation help selected constructed comparisons without resolving the primary-budget HMM targets. Width improvement is not uniform: the gate-only mean support-intersected width at $M=2400$ increases from 0.05202 to 0.05495. Both methods have zero observed family coverage failures; 100 replications give a zero-count Wilson upper endpoint of about 3.70\%, not proof of calibration. Appendix~\ref{app:eb-learned} separates crossings, terminal resolution and reused-data limitations.

\subsection{Scope of the guarantee}
Past-only refitting and predictable weights satisfy the sequential timing condition; retrospective selection on evaluation outcomes does not. The historical exercise remains exploratory, and declaring a family of 2,400 cannot account for an unrecorded wider search. Temporal-extension coverage is conditional on no evaluation feedback into specification choices; unknown external exposure and current-vintage data prevent claims of globally untouched confirmation or certified tradability. The guarantee concerns conditional capped losses, not economic utility or universal absence of predictive information.

\paragraph{Motivating reconstruction.}
The first reconstruction adds HML to a source-excluded SMB forecasting pipeline over 3,774 common 2010--2024 origins. Pooled, Gaussian and Student-$t$ incremental MSE gains are all negative, and their unrestricted forecasts trail zero return. Nine-lag, wider-penalty, annual-validation, symmetric-coding and eight-start sensitivities preserve negative HML inclusion gains, although some secondary model rankings change. Appendix details retain every result and distinguish exploratory sensitivity from the first configuration. These selected-pair findings motivate the complete-family evaluation rather than determine its expected sign.

\section{Frozen cross-market task family}
\subsection{Design, data and evidence status}
Before downloading the additional regional files or evaluating the temporal extension, we committed the expansion protocol to the revision branch (\texttt{84e0d36}). This is a timestamped repository freeze, not external preregistration. The family contains all 30 ordered distinct pairs of the six factors in the US, Europe, Japan, and Asia Pacific excluding Japan. It assesses three information-exclusion comparisons (pooled, Gaussian and Student-$t$) and two shared-gate ablations (Gaussian and Student-$t$). No direction or market is retained according to its performance.

The US panel contains 9,233 rows from 2 January 1990 through 31 August 2026. Each international panel has 9,348 rows from 1 November 1990 through 31 August 2026; its exact five-factor/momentum intersection excludes 88 earlier five-factor rows lacking momentum. Archive records determine the dates: the current regional momentum web description instead states a December 1990 start, a discrepancy retained in the provenance review. No observations are imputed or synchronized to the US calendar. The regional portfolios are arithmetic USD returns \cite{frenchdeveloped,frenchmom}; international archive headers identify the August 2026 Bloomberg database, while US headers identify CRSP. All files come from the French Data Library, and regions share global shocks and construction conventions. They are neither independent stochastic replications nor four independent data providers.

The fixed two-lag protocol uses pre-2005 training, expanding annual validation refits in 2005--2009, and annual evaluation refits in 2010--2026. The first trading observation of each calendar year starts a new block. HMMs use three states, eight starts and the preceding emission/optimization settings. All state blocks are included symmetrically with an unpenalized global intercept; standardized block coefficients receive equal ridge penalties. Validation chooses among $\{0.01,0.1,1,10,100\}$ and a full-history mean candidate separately for each path and target; the choice is frozen before 2010. This explicit mean candidate tests whether additional lag and gate flexibility earns its estimation cost. The same HMM fit can serve several targets without coupling their regression coefficients or validation choices. Eligible targets always differ from the excluded source.

We report 2010--2024 separately from January 2025--August 2026. The first period includes inspected US history and correlated external-market history; its confidence intervals are descriptive illustrations. The later period was outside prior repository forecast evaluation and current specification selection, although the original raw US archive already contained its observations. Exposure through the author's other work is unverified, so we do not certify globally untouched or independent confirmation. The sequential coverage assertion is conditional on the stated predictable-procedure and no-evaluation-feedback assumptions.

\subsection{Endpoints and multiplicity}
For each region and target the cap is $c=\max\{9\widehat{\operatorname{Var}}_{\rm pre2005}(Y),10^{-12}\}$, using population-denominator empirical variance. Ordinary MSE, RMSE, zero-return-normalized gains and cap activation remain descriptive endpoints. Capped losses address a different target. Each comparison is evaluated with unit weights and with each of the three source-excluded Student-HMM predictive weights, ordered by training covariance trace and held common across compared procedures. These are latent-state ranks rather than verified economic regimes. The simultaneous family size is $4\times30\times5\times4=2400$ for the temporal extension, with $\alpha=0.05$ and $\rho=1$ fixed in advance.

We retain all endpoints, first zero-crossing dates, interval widths and weight mass. Kish weight counts summarize concentration only; they are not dependence-adjusted effective sample sizes. An interval contained in $[-0.005,0.005]$ resolves a prespecified \emph{score-resolution} margin for the evaluated procedures. That margin has no established economic calibration and says nothing about oracle information, uncapped MSE or trading utility. Identical forecasts trivially have zero paired loss and an exact $[0,0]$ support bound; those cases must be separated from empirical equivalence of distinct forecasts.

\subsection{Measured family results and inferential resolution}
All four runs completed: 4,190 evaluation dates in the US and 4,347 in each international panel, with 1,232 selected HMM fits satisfying the implemented convergence criterion. No global-optimum claim follows. Independent code, importing none of the benchmark modeling functions, reconstructs 44,505 scalar forecasts across representative 2010 blocks with maximum absolute discrepancy $7.81\times10^{-17}$. It also checks every regional origin's annual cutoff, every saved penalty choice against validation losses, and every gate-fit input-prefix hash. This is independent implementation verification conditional on saved HMM parameters, not independent data replication or full EM re-estimation.

\begin{table}[ht]
\centering\small
\input{results/panel-comparison/family-table.tex}
\caption{Complete regional endpoint families. Positive/negative count terminal intervals excluding zero; no endpoint has an earlier crossing either. Margin counts intervals within $\pm0.005$ normalized capped loss. Historical rows are descriptive illustrations. All 776 temporal margin inclusions are implied by the predictable support bound alone; 196 are exactly identical mean forecasts. Counts overlap across factors and weights and are not independent replications.}\label{tab:panel-family}
\end{table}
There are 1,064 positive capped-loss point estimates among the 2,400 historical endpoints and 1,003 among the 2,400 temporal endpoints. Nevertheless, no endpoint in either period has a directional confidence-sequence crossing. Median full interval widths are 0.00641 and 0.01824, respectively, while median absolute point estimates are 0.0000623 and 0.000162. These unresolved directions cannot establish that the source lacks predictive information.

In the temporal family, 776 intervals lie within the declared margin: 196 pairs of identical selected-mean forecasts and 580 distinct but sufficiently close forecasts. Every one of these 776 inclusions is already guaranteed by $\sum_t w_tb_t/W_n\le0.005$. They therefore measure limited forecast separation; outcome-dependent concentration supplies no additional margin resolutions in this period. By contrast, 785 of the 1,650 historical margin inclusions follow from support alone, but historical selection precludes confirmatory interpretation of the remainder. Unweighted unrestricted cap activation ranges from 0 to 7.60\% in the temporal period (median 0.92\%), emphasizing that the score deliberately discounts some large errors.

\begin{figure}[H]
\centering
\includegraphics[width=0.98\linewidth]{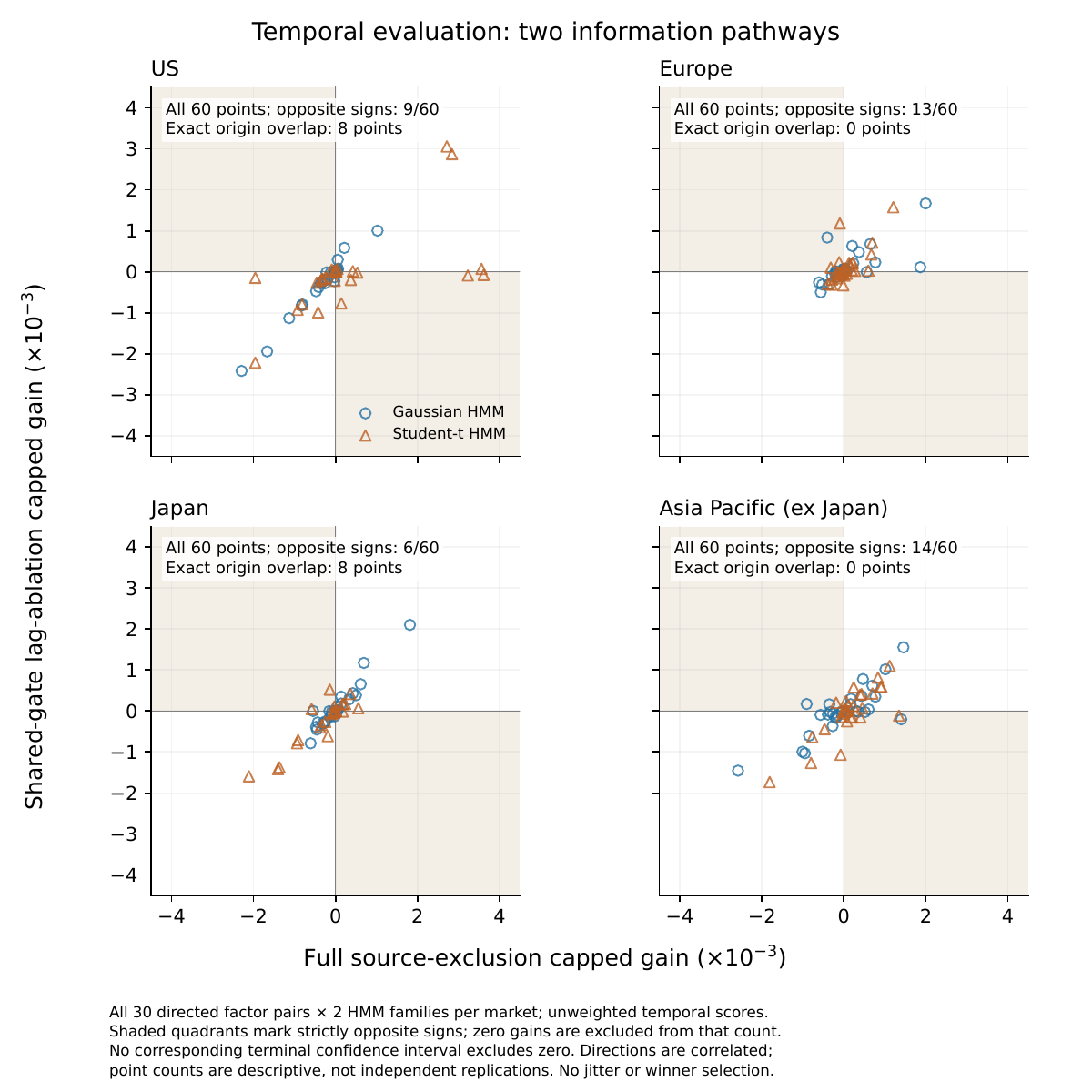}
\caption{Temporal unweighted capped-loss gains for every direction and both HMM families, comparing complete source exclusion with shared-gate lag ablation. Each region contains 60 comparisons. Opposite-sign quadrants contain 42 of 240 comparisons overall; exact zero cases are not counted as opposite signs. These point estimates are descriptive: none of their simultaneous intervals excludes zero.}\label{fig:pathway}
\end{figure}
Shared-gate and complete-exclusion point estimates have strictly opposite capped-loss signs in 50 of 240 unweighted historical comparisons and 42 of 240 unweighted temporal comparisons. Including the three predictable state weights gives 196 of 960 and 166 of 960, respectively. This is an observed sensitivity to the evaluated removal procedure, not a statistically established reversal of population predictive content. The differences may reflect representation, approximation, optimization and regularization as well as source information. All task-level magnitudes, zero cases and uncapped-loss comparisons are retained, avoiding selection of only disagreements.

\paragraph{Cap sensitivity and exact forecast-proximity support.}
A separately frozen post-review descriptive audit reuses every saved forecast and evaluates all 4,800 endpoints at caps of $4/9$, $1$, and $16/9$ times the primary cap (14,400 endpoint records). Relative to the primary cap, the smaller and larger caps change the signs of 357 and 136 of the 2,400 temporal gains. The corresponding complete-exclusion versus shared-gate sign-disagreement counts are 201, 166, and 163 of 960. Thus disagreement persists across these scores, but individual gain signs depend on how large errors are capped. Changing the cap changes the estimand; these are neither fresh confirmation nor additional significance tests.

For normalized capped squared loss, the exact outcome-uniform forecast-pair envelope is $b^\sharp=q(2-q)$, where $q=\min\{|r-u|/\sqrt c,1\}$. To see this, translate and scale the forecasts to $0$ and $\delta$. For $0\le\delta\le1$, the largest difference is attained when one squared error first reaches its cap, giving $1-(1-\delta)^2$; reflection gives the negative extremum. For $\delta\ge1$, the maximum is one. This elementary refinement uses only forecast separation and leaves the score unchanged. At the primary cap it reduces the temporal weighted support by a median 0.526\% among nonidentical pairs, increasing support-only margin inclusions from 776 to 777. It therefore offers little improvement in this application. The diagnostic computes deterministic support only and does not replace or select among the reported confidence sequences.

\paragraph{Magnitude of the procedure discrepancy.}
A post-review descriptive analysis directly computes $L(R)-L(\text{shared-}R)$ for every pair, verifying cancellation of the common unrestricted loss observation by observation. Among the 42 unweighted temporal sign reversals, both component capped gains are within $\pm0.005$. Their median absolute direct capped contrast is 0.000341 and the maximum is 0.003688. The corresponding absolute uncapped contrast normalized by zero-return MSE has median 0.173\% and maximum 4.22\%; capped and uncapped scales should not be conflated. Twenty-five reversals use the same validation-selected penalty in the two restricted models and 17 use different penalties; all 42 compare ridge predictors rather than mean fallbacks. Thus disagreement is not explained solely by mean selection or different penalty values.

The existing joint confidence event also yields a conservative interval for this contrast, $[l_R-u_S,u_R-l_S]$, by subtracting the two component intervals. Every such derived interval contains zero. This transformation needs no additional multiplicity allocation within the same covered family; it inherits all of its selection and temporal qualifications. The complete magnitude distributions and selection strata are retained. The analysis is explicitly post-review and does not replace the frozen endpoints or establish a meaningful conditional-risk reversal.

The original US HML-to-SMB direction has negative temporal point estimates under all five comparisons. Relative to zero-return MSE, the complete-exclusion gains are $-0.0867\%$ (pooled), $-0.1964\%$ (Gaussian), and $-0.2333\%$ (Student-$t$); shared-gate gains are $-0.1613\%$ and $-0.1415\%$. These descriptive losses extend the earlier unfavorable point-estimate pattern, but all bounded-score intervals contain zero. They neither validate an actionable warning signal nor prove universal nonpredictability.

A separate implementation recomputes all 4,800 terminal point estimates, bounds and decision flags, together with first-crossing dates, to maximum discrepancy $1.11\times10^{-16}$. The 5\% family guarantee is asserted conditionally for the temporal family only; displaying historical intervals does not create joint selection-adjusted coverage across both periods. A first crossing, if present, would concern the cumulative conditional target at that prefix, not permanent superiority.

\section{Controlled pathways and estimation costs}
\subsection{Known conditional means and matched procedures}
A separately frozen mechanism protocol evaluates four scenarios: null, direct-lag only, state-gate only, and mixed. The removed source is a \emph{block} $(S_t,L_t)$: $S_t\in\{-1,1\}$ is observed and flips with probability $q=0.05$, while $L_t$ is independent standard normal. The retained target and noisy control satisfy
\begin{align}
 C_t&=0.5S_t+\eta_t,\\
 Y_t&=0.2Y_{t-1}+\theta S_t+\beta L_{t-1}+\epsilon_t,
\end{align}
with independent standard-normal innovations. The four $(\theta,\beta)$ settings are $(0,0)$, $(0,0.8)$, $(1,0)$ and $(1,0.8)$. The label ``gate only'' describes this chosen architecture: a pooled model with the explicit state lag can represent the same state signal.

The exact unrestricted conditional mean is $m_{U,t}=0.2Y_{t-1}+\theta(1-2q)S_{t-1}+\beta L_{t-1}$. The exact restricted mean is $m_{R,t}=0.2Y_{t-1}+\theta\E[S_t\mid Y_{1:t-1},C_{1:t-1}]$, computed by a two-state filter receiving neither source coordinate. Its target residual emission has variance $1+\beta^2$, integrating the missing independent lag signal. The exact shared-gate mean keeps the state term in $m_U$ but omits $\beta L_{t-1}$. The chain starts with equal state probabilities, $Y_{-1}=0$, and an independent $L_{-1}$. The shared oracle conditions on retained history and the state-informed predictive gate while omitting the independent current source lag; this is a larger information set than the complete restriction. All oracles use past states or predictive probabilities, never the realized current state. Independent enumeration reproduces restricted weights within $2.22\times10^{-16}$, and independent conditional-mean Monte Carlo checks support the implementation.

Each of 100 independent seeds per scenario supplies 1,200 observations, with training through index 399, validation 400--599, and evaluation 600--1199; fits expand every 200 evaluation observations. Identical primitive seeds across scenarios support paired comparisons. Procedures include pooled ridge, known-parameter gated ridge, known-parameter gated least squares, and fitted Gaussian/Student-$t$ gates, with the declared ridge grid and mean fallback. The fitted HMMs use two states, four starts and at most 100 iterations. Seven of 6,400 selected HMM fits fail the convergence criterion and are retained. The pilot is stored separately and excluded from aggregates.

We evaluate both realized squared losses and the known conditional difference
\[
 \E[L(Y_t,r_t)-L(Y_t,u_t)\mid\mathcal F_{t-1}]
 =(r_t-m_{U,t})^2-(u_t-m_{U,t})^2.
\]
Averaging the latter removes realized innovation noise without removing uncertainty across simulated datasets. Reported Monte Carlo standard errors and approximate 95\% Student-$t$ intervals use independent seeds; they are not simultaneous tests or financial-data uncertainty estimates.

\subsection{Measured separation of the comparisons}
\begin{table}[ht]
\centering\small
\input{results/mechanism-full/mechanism-gains-table.tex}
\caption{Known-conditional squared-loss gains in the controlled experiment; mean across 100 seeds, with Monte Carlo standard error in parentheses. A shared-gate gain is not defined for the pooled comparator. The state/lag source block and exact filter are constructed controls, not empirical financial factors.}\label{tab:mechanism}
\end{table}
In the gate-only scenario, exact information value is 0.1301 (MCSE 0.0027), while exact shared-gate gain is zero. Fitted Gaussian and Student-$t$ information-exclusion gains are 0.1345 and 0.1374, but their shared-gate gains are both approximately $-0.0036$. Opposite fitted signs can therefore arise when the shared-gate oracle comparison is exactly zero: estimation costs should not be interpreted as negative information value. In the lag-only scenario both oracle gains coincide; in the mixed scenario information-exclusion gain is 0.8355 versus shared-gate gain 0.6457. The irrelevant-source null produces small negative fitted gains across procedures.

The pooled gate-only gain, 0.3061, exceeds the oracle information value because its restricted predictor approximates the retained-information conditional mean poorly. A larger fitted $R-U$ gap thus need not identify more oracle information or a better unrestricted forecast. This observation illustrates the earlier risk identity rather than establishing a new theorem.

Additional paired controls retain estimation costs separately from the gain contrasts. Replacing exact regression coefficients with known-gate least squares incurs positive average cost in every reported scenario/path; selected ridge-minus-OLS costs include regularization, validation tuning and the mean fallback, so they are not pure shrinkage effects. Replacing the known filter by a fitted HMM increases gate-only restricted risk by 0.0061 (Gaussian) and 0.0091 (Student-$t$). These differences include emission misspecification, optimization and downstream retuning; they cannot isolate pure gate-estimation error. The full cost tables and all seeds appear in the appendix and artifacts. This controlled result explains a possible procedural disagreement, but does not identify which mechanism generated any financial-data disagreement.

\subsection{Noisy-source failure study}
The observed-state design is favorable to gate recovery. A separately frozen stress study replaces its observed state coordinate by $X_t=S_t(-1)^{B_t}$, with independent $B_t\sim\operatorname{Bernoulli}(r)$ and $r\in\{0,0.1,0.3,0.5\}$, fixing the gate-only generator $(\theta,\beta)=(1,0)$. The removed block is now $(X,L)$; $L$ remains irrelevant. Neither exact filter receives the true latent state: the unrestricted filter combines past retained observations with the binary channel likelihood, and the restricted filter receives only $(Y,C)$. The exact shared-gate mean equals the unrestricted mean. At $r=0.5$, the channel is uninformative and both exact means coincide.

One hundred new seeds per corruption level give 400 runs, paired across levels using common primitives and a separate channel-noise stream. Sample sizes, fitting schedule and HMM budgets match the preceding experiment. All 6,400 selected HMM fits meet the convergence criterion; restricted fitted forecasts and selected penalties are exactly unchanged across each seed's corruption levels. No outcomes or noise levels are omitted.

Oracle information gains decline from 0.1336 at $r=0$ to 0.0708 at $r=0.1$, 0.0174 at $r=0.3$, and exactly zero at $r=0.5$. Oracle-gate ridge retains positive gains at the two intermediate levels. Fitted Gaussian and Student-$t$ gates instead lose: at $r=0.1$ their information-exclusion gains are $-0.0494$ (MCSE 0.0024) and $-0.0486$ (0.0023); at $r=0.3$ they are $-0.0216$ (0.0041) and $-0.0327$ (0.0044). Thus the favorable observed-state illustration does not imply robustness to a moderately corrupted source. At $r=0.5$, negative fitted gains coexist with exactly zero oracle information.

\begin{figure}[H]
\centering
\includegraphics[width=0.94\linewidth]{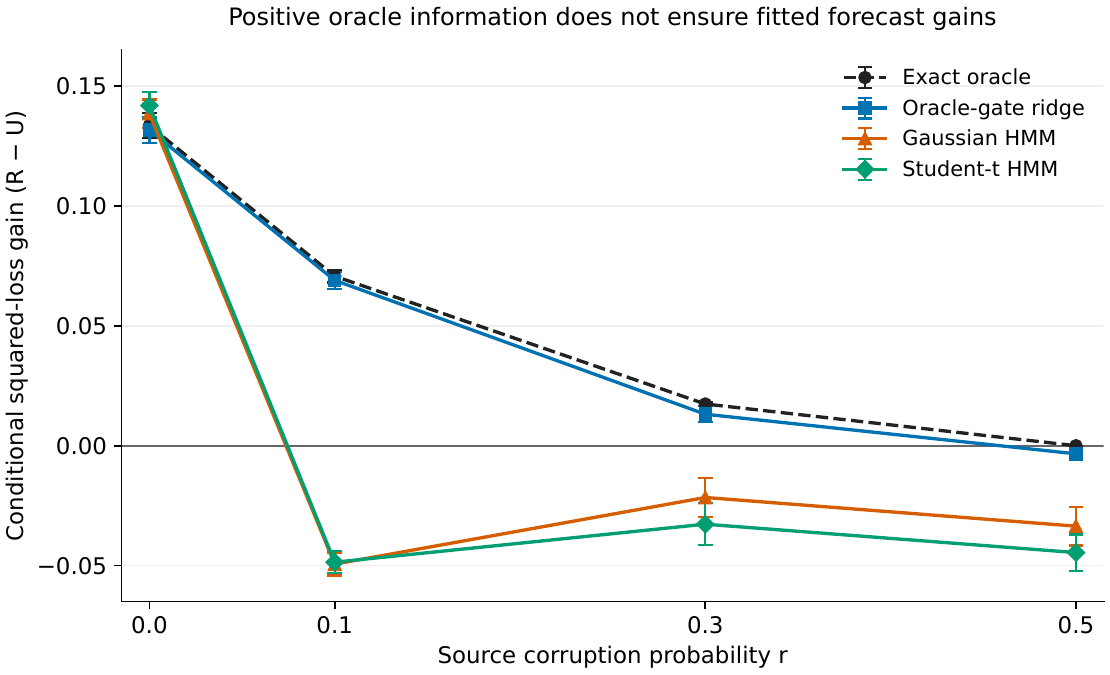}
\caption{Corrupting the source exposes fitted-gate failure despite positive oracle information. Points show mean known-conditional squared-loss gains in simulated target-squared units; bars are pointwise approximate 95\% Student-$t$ Monte Carlo intervals over 100 independent seeds per noise level. The same seeds are paired across levels, so points across levels are dependent. Lines connect the four evaluated settings only. Positive favors source inclusion. The exact shared-gate gain is zero at every level and is omitted. This is a controlled block-source experiment, not financial inference.}\label{fig:noisy-source}
\end{figure}

At $r=0.1$, unrestricted Gaussian/Student gate-replacement costs relative to oracle-gate ridge are 0.1250/0.1276, versus restricted costs 0.0066/0.0100. These combined approximation, estimation, optimization and retuning costs outweigh the oracle information advantage. They do not identify a unique failure mechanism. The matched observation-model intervention below tests a related explanation under shared privileged preprocessing; it does not identify the cause of failure in this original raw-input pipeline. Appendix~\ref{app:noisy-source} retains all gains, paired changes and costs. This block-source stress study closes part of the exact-state-observation limitation, not the financial external-validity gap.

\subsection{Matched observation-model intervention}\label{sec:emission-intervention}
A subsequent frozen experiment uses 100 fresh independent seeds, paired over the same four corruption levels, for 400 scenario-replications. Both fitted unrestricted gates receive $(E_t,C_t,X_t)$, where $E_t=Y_t-0.2Y_{t-1}$ uses the known generator coefficient, and omit the nuisance coordinate $L_t$. One arm treats all coordinates as conditionally independent Gaussian observations; the other uses Gaussian emissions for $(E,C)$ and a Bernoulli emission for $(X+1)/2$. The arms share initial hard memberships, four starts, the refitting schedule and regression candidate menu. The restricted gate and forecasts are identical across arms and corruption levels. State labels are ordered by fitted training-control means; latent states enter only held-out diagnostic scoring. All 3,600 unique selected fits meet the convergence criterion. This is established mixed-emission estimation, not a new estimator.

The primary contrast is the paired change in conditional unrestricted squared risk, mixed minus Gaussian, evaluated using the exact full-observed-past conditional mean. Predictive state probabilities are scored before the current observation is assimilated. Table~\ref{tab:emission-intervention} retains all four levels. At $r=0.1$, the mixed arm reduces conditional U risk by 0.12258, with a pointwise Monte Carlo interval for the signed contrast of $[-0.12831,-0.11686]$. Complete-exclusion gains change from $-0.05883$ (Gaussian) to $0.06375$ (mixed), compared with exact-oracle gain $0.06763$. Because R is shared, the gain improvement is exactly minus the U-risk contrast; these are not independent corroborating effects. The improvement in predictive Brier and logarithmic scores supplies a separate gate diagnostic.

\begin{table}[H]
\centering\scriptsize
\input{results/emission-intervention-full/emission-intervention-table.tex}
\caption{Matched mixed-minus-Gaussian contrasts. Negative risk, Brier and log-score differences favor mixed emissions. Brackets give pointwise approximate 95\% Student-$t$ Monte Carlo intervals over 100 independent seeds; the four levels use paired seeds, not 400 independent replicates. U risk uses simulated target-squared units. Floor fits count selected Gaussian refits with at least one proxy variance at the fixed standardized floor $10^{-4}$; denominators include validation and evaluation refits, not independent uncertainty units.}\label{tab:emission-intervention}
\end{table}

The effect is not uniform over corruption levels. At $r=0.3$, both arms have positive gains (0.01191 Gaussian and 0.01171 mixed), and the U-risk contrast interval includes zero despite small improvements in gate scores. The risk-contrast intervals also include zero at $r=0$ and $r=0.5$. Gaussian proxy variance floors occur in 394/400 selected refits at $r=0.1$, but also in every refit at $r=0$, where forecasts perform well. Floor contact alone therefore does not identify harmful forecasts. The Bernoulli probability floor $10^{-4}$ excludes exact boundary truth at $r=0$; that endpoint is a constrained approximation.

The intervention identifies a total procedure change within an oracle-residualized environment, including induced optimization and downstream tuning differences. Known-AR residualization and omission of $L$ are shared privileged changes relative to the preceding raw-input study, whose seed block also differs. In particular, the positive matched Gaussian gain at $r=0.3$ prevents attributing the earlier negative Gaussian result uniquely to its treatment of the binary channel. These observations support an observation-model explanation for the matched $r=0.1$ contrast, not a unique mediator, a finite-floor global optimum, uniform superiority, or a financial mechanism. No variance-floor sensitivity or feasible estimated-AR extension has been evaluated. Fit records, all-start diagnostics, forecast ledgers and all paired endpoint summaries are retained.

\section{Limitations and conclusion}
The revised evidence supports a procedural conclusion: source removal from a lag block and source removal from an entire estimated pipeline answer different questions. Their point estimates have opposite signs in 166 of 960 temporal comparisons, but the bounded-score intervals do not resolve any direction. The temporal margin resolutions follow entirely from forecast proximity. Consequently the cross-market expansion does not establish broadly useful predictive information, and failure to reject cannot establish oracle nonpredictability.

The four panels share a provider, portfolio conventions and global shocks; 30 directions share six series. Complete reporting prevents selection of only favorable tasks but does not create independent replications. The temporal extension improves separation from prior repository evaluation, while unknown outside exposure and retrospective vintages prevent certification of an untouched, real-time confirmation. Independent numerical replay verifies specified calculations, not external validity.

Financial model scope remains limited to pooled ridge and diagonal Gaussian/fixed-$\nu$ Student-$t$ gates with one validation era. Tail prediction, full-covariance models and other nonlinear architectures remain outside the measured claims. Capped-loss inference targets a bounded score and discounts sufficiently large errors; it does not give ordinary-MSE coverage or economic utility guarantees. The conservative simultaneous procedure fails to resolve known positive targets even in the learned-pipeline diagnostic; the matched-budget calibration does not establish financial power. The controlled simulations separate known predictive information from particular fitted-procedure costs, and the noisy-source extension demonstrates a failure of the fitted HMM procedures despite positive oracle information. The matched mixed-emission study identifies a narrower procedure effect under known-AR preprocessing. None of these studies establishes a financial mechanism.

The original HMM code, data vintage, search history and intermediate outputs remain unavailable. Its reported Granger statistics are therefore unreproduced, and a nine-lag specification still does not establish a nine-day warning horizon. The supported revision is a reproducible cross-market audit with explicit information paths and qualified inference, not a validated warning or trading system. Further confirmation requires documented evaluation isolation and evidence strong enough to resolve a scientifically meaningful target.

\appendix

\section{Linear hypotheses, identification and evaluation notes}\label{app:hypotheses}
These notes distinguish hypotheses and diagnostics from the executed fitted-procedure comparisons. Coefficient-equality tests and deployment evaluations discussed here are not reported as completed empirical tests.

\subsection{A precise linear hypothesis}
For a prespecified pair $(i,j)$, a baseline linear model is
\begin{equation}
 r_{j,t}=a+\sum_{\ell=1}^p b_\ell r_{j,t-\ell}
 +\sum_{\ell=1}^p c_\ell r_{i,t-\ell}+\varepsilon_t.
\end{equation}
The linear Granger null is $c_1=\cdots=c_p=0$. Rejection establishes a property of the conditional-mean specification, subject to model and inference assumptions; it does not establish unrestricted conditional dependence or structural causality. A test with $p=9$ concerns the \emph{joint contribution of lags 1 through 9}. It does not show that an effect appears exactly nine days later, lasts nine days, or gives nine days to hedge.

\subsection{Regime-coefficient equality}
To claim regime dependence, directly test equality of the relevant coefficient vectors across regimes, for example
\begin{equation}
 H_0^{\rm diff}:\ B_{j,\mathrm{crisis},\ell,i}
 =B_{j,\mathrm{normal},\ell,i},\qquad \ell=1,\ldots,p.
\end{equation}
A significant result in one regime and a nonsignificant result in another does not itself reject this equality. For nonlinear alternatives, test the corresponding difference in predictive contributions or held-out loss reductions.

\subsection{Identifying regime coefficients before testing them}
Let $x_t\in\R^d$ collect an intercept and the chosen lagged predictors, and abbreviate predictive weights as $w_{tk}$. Since $\sum_k w_{tk}=1$,
\begin{equation}
 x_t=\sum_{k=1}^K w_{tk}x_t.
\end{equation}
\begin{samepage}
Consequently a regression containing both $x_t$ and \emph{all} $K$ blocks $w_{tk}x_t$ is exactly rank deficient. One may instead use all $K$ gated blocks without a separate ordinary block, or use the reference-state form
\begin{equation}\label{eq:reference-gate}
 \widehat r_{j,t}=x_t^\top\beta_K+
 \sum_{k=1}^{K-1}w_{tk}x_t^\top\delta_k,
 \qquad \delta_k=\beta_k-\beta_K.
\end{equation}
This follows by replacing $w_{tK}$ with $1-\sum_{k<K}w_{tk}$. A difference from the reference regime is then a direct restriction on the relevant coordinates of $\delta_k$. For two nonreference regimes the restriction is on $\delta_k-\delta_m$.
\end{samepage}

Removing the algebraic redundancy does not guarantee identification. If the weights are constant across dates, only their weighted average coefficient vector is identified; the all-block design has rank at most $d$, not $Kd$. Nearly constant weights can create weak identification. Report design rank, singular-value condition number, state occupancy, and predictor scaling before interpreting coefficient contrasts. Reference coding also fixes a shrinkage geometry: equally penalizing base and contrast coefficients need not preserve forecasts when the reference state changes. We measure this sensitivity below. Ridge regularization can make predictions computable without creating likelihood-based identification of the unregularized regime coefficients. A fitted penalty does not justify ordinary unpenalized Wald $p$-values.

The included constructed example checks equivalence of the two parameterizations and the constant-weight failure case. It does not estimate an HMM or establish a regime effect in the factor data.

\subsection{Preserve the trading calendar}
Construct all lags on the complete ordered series before subsetting. Never concatenate separated crisis episodes and then shift that concatenated series. The earlier manuscript explicitly requires all lagged observations to remain in the same regime; this guards against crossing labels but changes the selected population. Retain this as a sensitivity analysis, not an unquestioned primary design. Equation~\eqref{eq:gated} retains transition observations and uses their actual lag history.

\subsection{Selection and inference}
Record every factor direction, regime, lag rule, emission family, and alternate specification inspected. A 30-pair correction addresses exactly that family, not all other selection decisions. If 30 directions are tested in three regimes, a simple 90-test Bonferroni reference threshold at family level $0.01$ is $0.01/90$. If multiple lags are separately tested and the smallest $p$-value is chosen, that search must also be accounted for. A BIC-based lag choice is a model-selection procedure; multiplying by the number of candidate lags is not a universal correction for its post-selection uncertainty.

Report conditional inference for a frozen specification separately from uncertainty in the full procedure. A full-procedure resampling or simulation analysis must repeat regime estimation, state alignment, and selection. For null-hypothesis calibration, generate under an explicit null and preserve relevant serial dependence and conditional heteroskedasticity. Resampling an unrestricted series without imposing the null does not automatically produce a valid null distribution.

\section{Earlier reconstruction procedures and deployment boundaries}\label{app:protocolnotes}
\subsection{Chronological development and evaluation}
Our implemented reconstruction uses an initial training period, a fixed chronological validation interval, and a later outer evaluation period. A single validation interval is simpler than nested rolling cross-validation and its representativeness is a limitation. In a walk-forward design, refit at a prespecified schedule using only observations available at that origin. Store an immutable prediction ledger with fit cutoff, data vintage, input features, predictive regime probabilities, selected hyperparameters, forecast, and realized outcome.

Because the published analysis already examined 1990--2024, reusing those years under a cleaner split is a \emph{pseudo-out-of-sample reconstruction}, not newly untouched confirmatory data. Freeze the revised specification before accessing a genuinely unused later period or an independent market. Describe that distinction explicitly. Do not tune repeatedly to the reconstructed outer test results.

\paragraph{The gate is a generated predictor.}
At each forecast origin, retain the gate's parameter-fit cutoff and its observation cutoff. A past-fitted HMM applied with smoothed probabilities still leaks future observations; a forward filter with full-sample-fitted parameters still leaks future parameter information. If a historical supervised training row is intended to mimic an online prediction, reconstruct its gate using the corresponding past-only fit or label that row as an in-sample fitting artifact. The reference helper checks neither of these timestamps; they must come from the experiment ledger.

\subsection{Forecast comparison}
Compare own-factor lags, all-factor linear prediction without regimes, Gaussian-gated prediction, Student-$t$-gated prediction, and a simple lagged-volatility regime model. Match tuning access and regularization. For a restricted baseline $R$ and an augmented model $U$, define the common-test-set loss difference
\begin{equation}
 d_t=L(r_{j,t},\widehat r^R_{j,t|t-1})
     -L(r_{j,t},\widehat r^U_{j,t|t-1}).
\end{equation}
Positive mean $d_t$ favors the augmented model. Report effect magnitude and dependence-aware intervals. Nested-model comparisons, estimated gates, and repeated selection may invalidate a naive independent-sample test, so the inference procedure must be justified for this design.

For a horizon claim, separately estimate direct $h$-step models and evaluate prespecified $h$, rather than reinterpreting lag order as horizon. Report one-day prediction as the primary endpoint and any multi-horizon search as a declared family. Regime-stratified results should use the ex ante regime probabilities or training-fixed alarm rule, not labels selected using the realized test outcome.

\subsection{Implemented forecast procedures}
The pooled baseline and the own-target-lag baseline use ridge regression. A simple gate uses only target history, so it is admissible in both source-excluded and source-included models:
\begin{equation}
 q_t=\frac{\sqrt{5^{-1}\sum_{s=1}^{5}r_{j,t-s}^2}}
 {\sqrt{20^{-1}\sum_{s=1}^{20}r_{j,t-s}^2}},\qquad
 w_t=\{1+\exp[-4(q_t-1.5)]\}^{-1}.
\end{equation}
Set $q_t=1$ for an all-zero history. These are engineered soft weights, not calibrated crisis probabilities. The first $\max(p,20)$ observations are excluded from supervised fitting as warmup.

The estimated-gate comparisons use diagonal-scale Gaussian and Student-$t$ HMMs. The latter fixes $\nu=5$; it does not estimate state-specific tail parameters. For an information-exclusion comparison, the restricted HMM fits and filters all retained factors except HML, while the unrestricted HMM may use all six. Each gate feeds a separate ridge predictor using the corresponding factor lags. Gaussian and Student-$t$ models have the same state count, optimization budget, and forecast split. They differ in emission distribution, not in their access to future observations.

At each refit, all HMM scaling and parameters are fitted on rows preceding the block. A fixed-parameter forward filter then provides one-step predictive weights through that block. Historical features within the current supervised fit sample use the HMM fitted on that entire fit sample. They are explicitly \emph{in-sample generated features}, not historical online forecasts. No validation or evaluation observation enters the preceding parameter fit; observations already revealed within a block may update its filter. Gate and regression parameters are both refitted at each scheduled cutoff.

For each regression, standardize design columns on its fit sample and minimize mean squared error plus $\alpha\|\beta\|_2^2$. The ordinary intercept is unpenalized; standardized lag and gate-contrast columns are penalized. Select $\alpha$ separately by validation MSE, with ties resolved by the first grid entry, and hold it fixed during outer evaluation. Scaling is recomputed at each refit. Zero-return and expanding historical-mean predictions provide low-complexity reference forecasts. No HMM state is declared a verified economic ``crisis'' on the basis of its latent label.

\subsection{Risk-management value}
If the aim is monitoring rather than trading, define the actual target: future tail exceedance, drawdown, volatility, or an exposure-loss function. Compare with a simple volatility-only forecast. For alarms, prespecify onset criteria, threshold, persistence, matching window, and counting rules. Report false alarms per year, missed events, time in alarm, and detection delay. Lead time without these denominators is incomplete.

A hedge or exposure-control rule requires a specified loss or utility, execution timing, transaction costs, and a benchmark with the same risk budget. Predictive significance alone does not establish hedging benefit. Any common-driver or forced-liquidation interpretation remains a mechanism hypothesis unless independently tested with holdings, liquidity, or flow data.

\section{Sequential boundary proof and calibration details}\label{app:sequential}
The following proof establishes the normal-mixture proposition in the main text under its stated assumptions.
\begin{proof}
For one comparison write $S_n=\sum_{t\le n}w_t(d_t-\mu_t)$. Conditional
Hoeffding's lemma, using the support interval $[-w_tb_t,w_tb_t]$ for the uncentered variable $w_td_t$ (centering translates this interval without changing its width), gives
\[
 \E[\exp\{\lambda w_t(d_t-\mu_t)\}\mid\mathcal F_{t-1}]
 \le\exp\{\lambda^2(w_tb_t)^2/2\}
\]
for every real $\lambda$. Thus $\exp\{\lambda S_n-\lambda^2V_n/2\}$ is a
nonnegative supermartingale. Mixing $\lambda$ against $N(0,1/\rho)$ and using
conditional Tonelli yields
\[
 \mathcal M_n=\sqrt{\frac{\rho}{\rho+V_n}}
       \exp\!\left\{\frac{S_n^2}{2(\rho+V_n)}\right\},
 \qquad \mathcal M_0=1.
\]
Ville's inequality bounds the probability of ever having
$\mathcal M_n\ge1/\alpha_j$ by $\alpha_j$. Solving this inequality gives
$|S_n|\le u_j(V_n)$ simultaneously over time outside that event. Division by
$W_n$ and a union bound over comparisons prove the claim.
\end{proof}

\subsection{Executed calibration and power limitations}
We evaluated the implementation on 12 synthetic scenarios: four environments
(independent, persistent, abrupt-change, and adaptive-variance) crossed with
null, weak, and strong alternatives. Each scenario used 2,000 independent
replications, 500 time steps, and 90 correlated comparisons, for 24,000
scenario-replications. The nominal family error was $0.05$, with $\rho=1$
and seed 20260928. Two-point outcomes give exactly computable conditional
loss differences, so coverage can be checked against the changing target.
The 90-comparison calibration does \emph{not} validate the power of the
2,400-comparison application.

No scenario showed a family coverage failure through the 500 evaluated steps. For each scenario,
zero failures in 2,000 replications has a two-sided 95\% Wilson upper endpoint
of approximately $0.1917\%$. This is finite Monte Carlo evidence under the
constructed distributions, not proof of zero failure probability or exact
nominal calibration. Positive-boundary crossing was zero under every weak
alternative. Under strong alternatives it was 99.4\% in the independent
environment, 69.6\% under adaptive variance, and zero in the persistent and
abrupt-change environments. Positive crossing is a type-I event in null rows
and a power measure in alternative rows; it is not the coverage-failure
event. Mean final \emph{untruncated} interval widths ranged approximately
from 0.188 to 0.206 in normalized capped-loss units. These results demonstrate
conservatism and potentially severe power limitations rather than uniformly
effective detection. Clipping is active in several scenarios, but their
bounded two-point outcomes do not establish efficacy for heavy-tailed
financial returns.

\subsection{Selection, provenance, and scope of inference}
Past-only model refitting and weight adaptation are compatible with the
theorem. Retrospectively selecting the factor pair, model family, loss cap,
or boundary tuning after inspecting evaluation outcomes is not. Merely
declaring a family size of 2,400 does not account for an unrecorded wider
search. The historical 1990--2024 exercise therefore remains exploratory;
the confidence-sequence machinery cannot convert previously studied data
into new confirmatory evidence.

A later temporal extension must retain its actual provenance. A period can
be new to this execution while prior exposure by the author or another
research session remains unknown. Such an extension should not be described
as genuinely untouched without supporting records. Current-vintage revised
factor observations also do not establish their real-time availability at
the historical forecast origin. The theorem addresses sequential
conditional loss under its stated filtration, not the certification of a
tradable information set. The mathematical guarantee survives arbitrary
temporal dependence under its assumptions, but makes no promise of future
performance, economic utility, causal interpretation, or universal absence
of predictive information.

\section{Stopped-range empirical-Bernstein adaptation}\label{app:variance-adaptive}
This diagnostic was specified after inspecting the primary financial and learned-pipeline results. Its code and fresh synthetic comparison were frozen in commit \texttt{173f800} before their outcomes. It retains the capped-loss target, predictable weights and forecast-pair bounds defined in the main text. The following argument adapts established exponential-supermartingale constructions \cite{howard2021,choe2024}; it does not replace the primary analysis or assert methodological novelty.

Write $D_t=w_td_t$, $a_t=w_tb_t$, $Q_n=\sum_{t\le n}D_t^2$, $A_n=\max_{t\le n}a_t$ and $W_n=\sum_{t\le n}w_t$. Fix $J=11$ ranges $B_j\in\{2^{-10},2^{-9},\ldots,1\}$. For each range, let $\tau_j=\inf\{t:a_t>B_j\}$: its process stops permanently \emph{before} the first violating observation is incorporated. The stopping decision is predictable. Set
\[
 \lambda_m=0.9\,2^{-m},\qquad
 \pi_m=\frac{1}{(m+1)(m+2)},\qquad
 \psi(\lambda)=-\log(1-\lambda)-\lambda,
 \quad m=0,1,\ldots.
\]
The weights $\pi_m$ sum to one. For $W_n>0$, define
\begin{equation}\label{eq:eb-radius}
 r_n=\min_{\substack{j:B_j\ge A_n\\0\le m\le K(n)}}
 \frac{B_j\log\{2MJ/(\alpha\pi_m)\}
       +\psi(\lambda_m)Q_n/B_j}{\lambda_m W_n},
 \qquad K(n)=\lceil\log_2(n+1)\rceil+2.
\end{equation}
The resulting interval centers on $\sum D_t/W_n$ and is intersected with $[-\sum a_t/W_n,\sum a_t/W_n]$. At zero weight mass the target is undefined. If all predictable bounds are zero, the support intersection is exactly $[0,0]$. In contrast, observed $Q_n=0$ with nonzero predictable bounds does not justify a zero radius. The implementation records radius zero when $A_n=0$ using the exact support certificate; in that special case the stored radius is a support shortcut rather than the literal untruncated expression in~\eqref{eq:eb-radius}.

\paragraph{Simultaneous guarantee under the stated assumptions.}
For $x\in[-1,1]$ and $0<\lambda<1$, the power series for the logarithm gives
\[
 \lambda x-\log(1+\lambda x)\le\psi(\lambda)x^2,
 \qquad e^{\lambda x-\psi(\lambda)x^2}\le1+\lambda x.
\]
Consequently, with $X_{j,t}=\mathbf1\{t<\tau_j\}D_t/B_j$ and $\nu_{j,t}=\E[X_{j,t}\mid\mathcal F_{t-1}]$,
\[
 \E[e^{\lambda(X_{j,t}-\nu_{j,t})-\psi(\lambda)X_{j,t}^2}
       \mid\mathcal F_{t-1}]
 \le e^{-\lambda\nu_{j,t}}(1+\lambda\nu_{j,t})\le1.
\]
Products of these factors, and their negative-tail counterparts, are nonnegative supermartingales starting at one. Apply Ville's inequality with error $\alpha\pi_m/(2MJ)$, then union over comparisons, both tails, ranges and the countable tuning family. Whenever $B_j\ge A_n$, no increment has been stopped through $n$, so the stopped sum is the full centered weighted difference divided by $B_j$. Every eligible radius in~\eqref{eq:eb-radius} therefore covers the same target on one simultaneous event. Minimizing over eligible ranges and the deterministic finite prefix of tuning indices preserves coverage, jointly over all times and the prespecified family. The countable allocation, rather than a fixed positive minimum tuning parameter, justifies the growing computational prefix. No particular asymptotic rate or optimality is claimed.

\paragraph{Implementation and validation scope.}
For each tuning index the numerator has form $uB+v/B$. Its exact minimum on eligible dyadic ranges lies at one of the two neighboring ranges of its constrained continuous minimum, or at an endpoint. An independent scalar implementation checks this optimization, chronological prefix invariance, range equality and zero-observation edge cases. The quantity $Q_n$ is an observed raw second moment; it is not an estimated conditional variance substituted into a Hoeffding bound. All ranges and tuning indices share an explicit error budget. This within-method correction does not cover post hoc selection between this method and the original normal-mixture procedure.

\subsection{Paired learned-pipeline comparison}\label{app:eb-learned}
The learned comparison was frozen in commit \texttt{93e52e3} after the old MSE and normal-mixture results were inspected. It reuses all 400 mechanism ledgers: no fresh seeds, HMM fits or independent confirmation are added. Conditional Gaussian-mixture capped risks, training-random caps, and predictable exact restricted-state weights are unchanged. All 26,400 paired endpoint records reproduce the earlier normal-mixture terminal means, bounds and targets within $10^{-14}$, with identical coverage flags. Each dataset supplies 33 actual endpoints and 600 prefixes; $M=2400$ is an allocation charge, not 2,400 distinct learned procedures.

\begin{table}[htbp]
\centering\small
\input{results/variance-adaptive-learned/paired-learned-table.tex}
\caption{Paired normal-mixture (N) and empirical-Bernstein (E) diagnostics on reused learned forecasts. Fail, Pos and Neg count datasets with any coverage failure, positive crossing or negative crossing across 33 endpoints and all prefixes, out of 100 seeds per scenario. Width averages support-intersected terminal widths over endpoints and seeds. Each allocation and method is reported separately; choosing whichever rejects has no joint 95\% guarantee.}\label{tab:eb-learned}
\end{table}
At $M=2400$, the three positive datasets in the mixed scenario contain pooled complete-exclusion crossings only; neither fitted HMM has a positive crossing. At $M=33$, unweighted Gaussian information-exclusion comparisons cross positively in 43 lag-only and 45 mixed datasets; the corresponding Student-$t$ counts are 43 and 48. These are \emph{anytime} counts. Terminal positive resolution counts are smaller: Gaussian 42/42 and Student-$t$ 42/45 for lag-only/mixed, respectively. Neither method resolves a gate-only direction. Crossing at an earlier prefix does not imply terminal or permanent superiority.

There are no observed coverage failures, negative crossings or erroneous directions in these runs. Zero failures in 100 independent seeds has a pointwise 95\% Wilson upper endpoint of approximately 3.70\%; endpoints, weights and shared-seed scenarios are not extra independent replicates. Realized positive-target subsets describe conditional resolution, not unconditional power populations. Exact mean-oracle forecasts are not asserted to minimize capped loss, and fitted forecasts in the null generator need not have equal risk. The null exact-oracle forecasts coincide, so their type-I control is intentionally trivial.

Mean supported widths improve in most rows but worsen from 0.05202 to 0.05495 in gate-only at the primary allocation. Raw-width summaries additionally require care: when the entire predictable support is exactly zero, the empirical-Bernstein implementation records zero radius from the support certificate, not from the literal exponential-bound formula. Such reductions cannot be attributed to learning empirical variance. For nonzero predictable support, zero observed squared differences do not erase the logarithmic penalty. This exploratory paired diagnostic neither validates financial forecast power nor permits retrospective best-method selection.

\section{Vintage and publication-availability notes}\label{app:availability}
Choose $K$, Gaussian versus Student-$t$ emissions, and tuning parameters within chronological development data. Map latent labels to descriptive names using a training-only rule, such as increasing covariance trace. This avoids reassigning ``crisis'' to whichever state best matches known test events. Heavy tails do not guarantee superior crisis detection; compare predictive scores, alarm behavior, and calibration under the same information constraints.

\paragraph{Vintage and publication availability.}
The French Data Library switched its US research-return production from CRSP FIZ to CIZ beginning with the January 2025 release. The documented monthly-return construction also changed: CIZ compounds daily returns and reinvests dividends on ex-dates rather than month-end \cite{french}. The replication manifest must therefore record the file hash, download date, return convention, and vintage; the switch cannot be inferred merely from a return observation's calendar date.

The research clock and publication clock are distinct. A dated historical factor observation is not proof that the published series was available immediately after that day's or month's close. Unless archived release timestamps or an implementable reconstruction establish availability, a lagged exercise is a chronological research backtest, not a certified real-time trading backtest. Use $\mathcal F_t^{\rm avail}$ for genuinely available inputs and record observation date, release time, and execution time separately.

\section{Current-vintage historical reconstruction}
\subsection{Design and provenance}
The reconstruction forecasts SMB from a source-excluded set containing MKT-RF, SMB, RMW, CMA and MOM, and a source-included set adding HML. Initial HMM and regression fits use 1990--2004 (3,784 rows); 2005--2009 supplies 1,259 validation rows. Evaluation covers 3,774 common origins from 4 January 2010 to 31 December 2024. This is pseudo-out-of-sample reconstruction of previously studied years, not untouched confirmation.

The primary configuration has two lags; nine lags are a declared sensitivity tied to the earlier manuscript. Both refit every 252 observations, use three HMM states, two starts, and at most 200 iterations. Every selected HMM fit converges in both runs (128 fits total), although convergence to a global optimum is not established. Data, protocol, source-code hashes, daily predictions and fit diagnostics are retained. The configuration was recorded before inspecting these empirical losses; it is not externally preregistered.

\subsection{Forecast errors and incremental value}
\begin{table}[ht]
\centering\small
\input{results/empirical-comparison/rmse-table.tex}
\caption{SMB root mean squared forecast error in basis points, on the same 3,774 evaluation dates. Lower is better. These are new measured current-vintage results, not the original preprint's statistics. The zero-return benchmark has the lowest aggregate error among the displayed procedures.}\label{tab:empirical}
\end{table}
For two lags, the source-exclusion MSE differences are $-9.11\times10^{-8}$ (pooled), $-1.09\times10^{-7}$ (Gaussian HMM), and $-8.41\times10^{-8}$ (Student-$t$ HMM), in decimal-return squared units. All favor the source-excluded procedure. The unrestricted pooled, Gaussian, and Student-$t$ forecasts increase MSE relative to zero return by approximately 0.94\%, 1.43\%, and 1.18\%, respectively. At nine lags, every source-exclusion comparison remains negative, including the simple volatility gate.

\begin{figure}[H]
\centering
\includegraphics[width=0.96\linewidth]{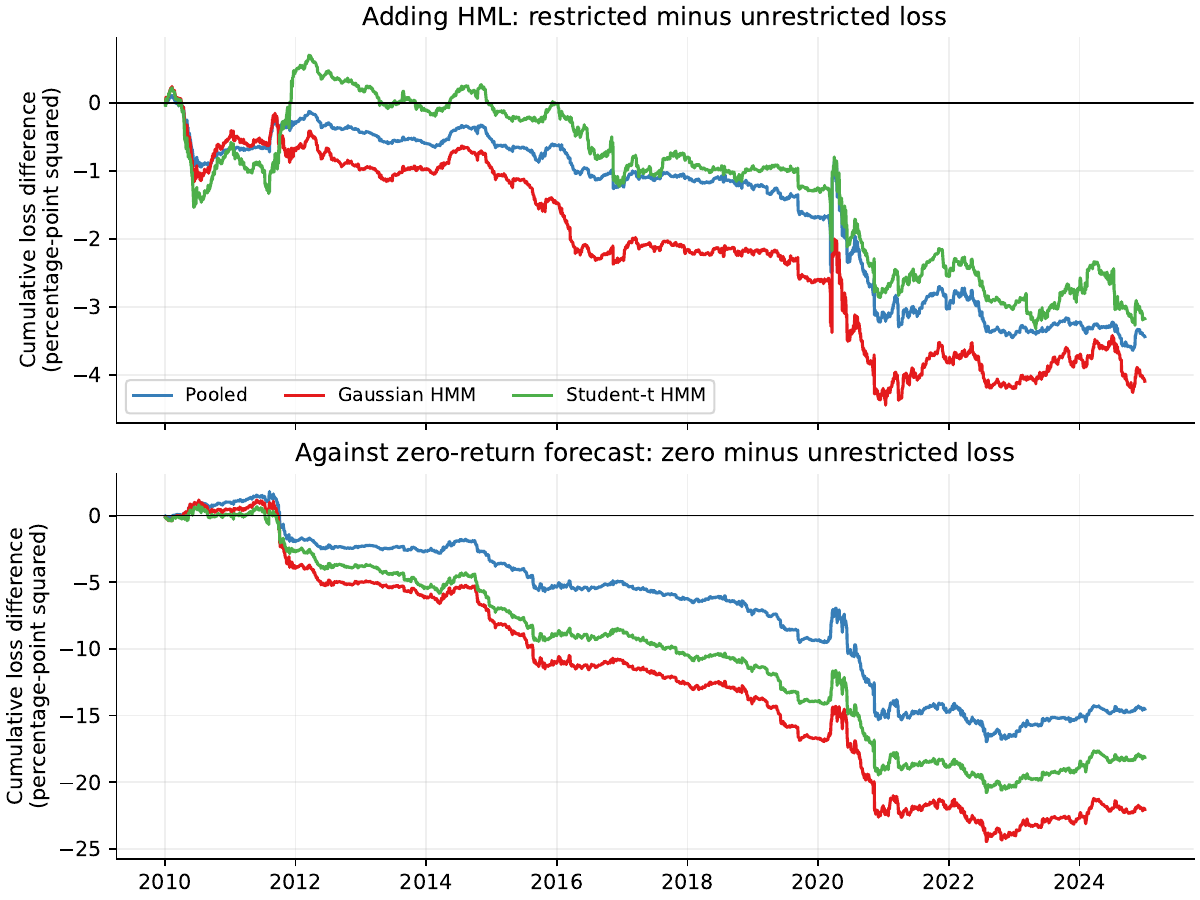}
\caption{Primary two-lag cumulative paired losses. Positive values favor source augmentation (top) or the augmented forecast over zero return (bottom). These curves summarize realized errors, not portfolio wealth or trading returns.}\label{fig:cumulative}
\end{figure}
Performance varies over time, so an aggregate negative gain does not rule out selected positive episodes. Yearly comparisons are released without choosing favorable periods as confirmation. The cumulative curves likewise do not establish a change point or a causal effect.

\subsection{Exploratory ex ante regime weighting}
An all-period average can conceal state-specific performance. As a disclosed analysis added after inspecting aggregate results, we weight each common loss difference by the predictive probabilities of the \emph{source-excluded Student-$t$ gate}, using the same reference weights for every model:
\begin{equation}
 \widehat\Delta_{s,m}=\frac{\sum_t w^{R,\mathrm{St}}_{t|t-1,s}d_{t,m}}
 {\sum_t w^{R,\mathrm{St}}_{t|t-1,s}}.
\end{equation}
The superscript $R,\mathrm{St}$ denotes the source-excluded Student-$t$ model, not a realized-state label. At each past-only fit, state ranks are ordered by covariance trace. This does not establish invariant economic state identities across refits. No evaluation outcome or smoothed probability is used to assign its evaluation weight.
\begin{table}[ht]
\centering\small
\input{results/empirical-comparison/regime-table.tex}
\caption{Two-lag source-exclusion gains weighted by the reference gate; gains are in $10^{-8}$ decimal-return squared units. Weight mass is the sum of probabilities, not a count of identified crises or an effective sample size. Positive favors source augmentation.}\label{tab:weighted}
\end{table}
The high-volatility-rank weighted gain is negative for all three predictors and remains negative in the nine-lag and extended-grid diagnostics. The low-rank Gaussian gain is positive in the primary run, illustrating why the conclusion is about aggregate performance, not absence of every local association. These post-primary weighted summaries have no calibrated regime-difference test and cannot verify or refute the original crisis-regime coefficient claim.

\subsection{Uncertainty and regularization sensitivity}
We resample paired loss vectors with the stationary bootstrap \cite{politis}, using 2,000 draws, seed 20260926, and mean block lengths 5, 20 and 60. Reported 2.5th--97.5th percentile ranges are conditional on recorded forecast paths: gates, model selection and regressions are not refitted within these draws. Stationarity and the conditions needed for valid confidence coverage are not established for this expanding-fit series. The ranges are pointwise descriptive sensitivity diagnostics, not calibrated noncausality tests or simultaneous confidence intervals.

At mean block length 20, the two-lag Student-$t$ source-exclusion range is approximately $[-2.12,0.38]\times10^{-7}$. It includes zero. The pooled range is approximately $[-1.91,0.04]\times10^{-7}$; whether its upper endpoint exceeds zero changes across block lengths. We therefore make no significance claim based on selecting a favorable block length. Neither a negative point estimate nor these ranges establish population absence of predictive information.

All primary models selected $\alpha=1$, the largest value in the original validation grid. After observing this boundary behavior, we recorded an explicit protocol amendment and ran an exploratory two-lag sensitivity extending the grid to 10 and 100, using the same pretest validation years. The primary results are retained. Source-augmentation gains remain negative for every model in this sensitivity, and all augmented models still trail the zero-return benchmark. Errors become much closer to zero-return performance, showing that regularization materially affects the magnitude of the apparent disadvantage. This additional run cannot be presented as a separately preregistered confirmation.

\subsection{Exploratory specification checks}
These checks were designed after reviewing the primary outcomes and are exploratory. First, we align penalty selection with deployment: five expanding validation blocks refit every 252 observations during 2005--2009, using the extended grid and the same 2010--2024 evaluation dates. Penalty values use only pre-2010 validation losses, while the sensitivity design was motivated by previously inspected evaluation results. Source-exclusion gains remain negative: $-1.48$, $-1.31$, $-7.48$, and $-1.50$ in $10^{-8}$ squared decimal-return units for pooled, volatility-gated, Gaussian, and Student-$t$ procedures. All augmented forecasts still trail zero return. This removes a mismatch in fitting schedules; it does not remove historical hypothesis selection or establish population noncausality.

Second, reference coding preserves the unpenalized regression span but not an isotropic ridge penalty on base and contrast coefficients. We reuse the archived HMM fits, vary the reference among all three volatility ranks, and retune penalties on the original pretest validation interval. We also evaluate a symmetric encoding with all $K$ gated feature blocks, a free global intercept, and a common ridge penalty on standardized columns (all selected symmetric penalties are positive). Under a state permutation these columns are merely permuted: their training means/scales and coefficient penalty permute together, so predictions are invariant. The redundant design is acceptable for this positive-penalty prediction calculation; it does not identify unpenalized state coefficients. This invariance is a standard algebraic property, not a new estimator claim.

\begin{table}[ht]
\centering\small
\input{results/empirical-comparison/audit-sensitivity-table.tex}
\caption{Post-audit two-lag sensitivities, in $10^{-8}$ squared decimal-return units. Exclusion compares independently source-excluded and source-included gates. Shared-gate compares source-lag exclusion versus inclusion while retaining the source-informed gate; it is not an information-exclusion test. All penalties are selected before test evaluation from the extended grid.}\label{tab:audit-sensitivity}
\end{table}
The magnitude changes with reference choice, but both gains remain negative in every row. Every augmented model also has higher MSE than zero return. Replaying the primary fits with the original penalty reproduces all HMM forecasts exactly, isolating regression coding/tuning from changes in HMM estimation. The shared-gate diagnostic distinguishes the explicit-lag pathway from complete source exclusion without making an additive decomposition of the two fitted gains.

We additionally release high-minus-low weighted gain contrasts, per-year state-weight mass and Kish weight-concentration counts. Approximately 68\% of high-rank probability mass falls in 2020--2022, including 27.7\% in 2022 alone. Period concentration and changing fitted gates limit interpretation as invariant economic states. These counts are not dependence-adjusted effective sample sizes; neither the contrasts nor the weighting establish a latent-state causal effect or a calibrated regime-difference test.

\paragraph{Optimization sensitivity.}
An additional post-audit run increases HMM initialization starts from two to eight, retaining three states, fixed degrees of freedom, the original fitting schedule and the extended pretest penalty grid. All 64 selected fits converge. Gaussian and Student-$t$ source-exclusion gains remain negative, at $-1.61\times10^{-9}$ and $-4.17\times10^{-9}$ squared decimal-return units, respectively; both augmented forecasts remain worse than zero return. This is optimization sensitivity with validation retuning: the Student-$t$ penalty changes from 10 to 100. Its comparison with unrestricted pooled forecasting reverses in favor of the HMM, although source inclusion still loses and zero return remains better. Optimization changes magnitudes and secondary rankings, so emission-family differences should not be treated as structural facts. Eight starts reduce one diagnostic concern without certifying a global optimum.

\section{Repeated simulation diagnostics}
We distinguish algorithm checks from financial evidence. Four constructed processes each contain a source $X$, target $Y$, and observed control $C$, with 1,200 observations and two forecast lags. Training indices are $0$--$399$, validation is $400$--$599$, and evaluation is $600$--$1199$, with refits every 200 observations. The declared ridge grid is $\{0,0.01,0.1,1\}$. Independent replications use seeds 1000 onward, retaining every outcome.

The null process is $Y_t=0.2Y_{t-1}+\epsilon_t$, with independent standard-normal source, control and innovations. The linear alternative adds $0.8X_{t-1}$. The heavy-tailed null uses independent $t_3/\sqrt{3}$ innovations and factors; finite variance but infinite fourth moment makes squared-loss uncertainty particularly unstable. In the switching alternative, a stationary two-state chain flips with probability $0.03$ and
\begin{align}
 C_t&=2(2z_t-1)+\eta_t,\\
 Y_t&=0.2Y_{t-1}+0.8(2z_t-1)X_{t-1}+(0.5+1.5z_t)\epsilon_t.
\end{align}
Here $X_t$, $\eta_t$, and $\epsilon_t$ are independent standard normal and independent of the chain. The control is a deliberately informative noisy proxy; the latent state itself is never supplied to a forecaster. This favorable constructed alternative is a recovery check, not an external-validity argument.

We execute 100 replications per process for the simple baselines and 20 per process for fitted two-state HMMs, for 480 runs in total. HMM runs use two starts and at most 200 EM iterations. Of 1,280 selected HMM fits across the repeated study, one fails the convergence criterion; it is retained and flagged. The smaller pilot used for timing is stored separately and excluded from these aggregates.

\begin{table}[ht]
\centering\small
\input{results/empirical-comparison/simulation-table.tex}
\caption{Mean source-exclusion gain $R-U$ in the 20-replication HMM study; parentheses give Monte Carlo standard errors across independent replications. Units are synthetic squared-return units. These are not standard errors for the financial data or null-test rejection rates. The heavy-tailed row reports finite-run dispersion without a finite-variance Monte Carlo error guarantee.}\label{tab:simulation}
\end{table}
Pooled forecasts recover the linear signal, whereas HMM gates recover the switching signal more effectively than the pooled model. No tested gate has a positive average source-augmentation gain under either null in this study. The Student-$t$ gate is not uniformly better than the Gaussian gate: under the Gaussian switching generator its average unrestricted loss is higher. Extra complexity is therefore not treated as an automatic improvement.

\section{Audit of the reported historical evidence}
\subsection{Multiple-testing arithmetic}
Table~\ref{tab:pvalues} applies the original manuscript's stated threshold to its reported $p$-values. These are arithmetic corrections, not recalculated statistical tests.
\begin{table}[ht]
\centering\small
\begin{tabular}{llrr}\toprule
Regime & Direction & Reported $p$ & Below $0.01/30$?\\\midrule
Normal & HML $\to$ SMB & $1.52\times10^{-2}$ & No\\
Normal & SMB $\to$ HML & $9.81\times10^{-2}$ & No\\
Elevated & HML $\to$ SMB & $8.70\times10^{-2}$ & No\\
Elevated & SMB $\to$ HML & $1.94\times10^{-4}$ & Yes\\
Crisis & HML $\to$ SMB & $1.89\times10^{-5}$ & Yes\\
Crisis & SMB $\to$ HML & $1.65\times10^{-1}$ & No\\\bottomrule
\end{tabular}
\caption{Original Table 3 values with corrected classification under the explicitly stated 30-test family. The family may be too narrow for the actual selection process.}\label{tab:pvalues}
\end{table}
Since $0.01/30\approx3.33\times10^{-4}$, the elevated reverse-direction result passes that threshold. Under a 90-test family the threshold is approximately $1.11\times10^{-4}$, so only the crisis direction passes among these rows. The correct family must follow the actual analysis history; it cannot be chosen after viewing which result survives.

\subsection{Event-level consistency}
The original event table defines success as $p_{HML\to SMB}<0.10$ \emph{and} $p_{SMB\to HML}>0.10$. Table~\ref{tab:events} applies that rule consistently.
\begin{table}[ht]
\centering\small
\begin{tabular}{lrrl}\toprule
Episode & Forward $p$ & Reverse $p$ & Stated rule\\\midrule
2008 financial crisis & 0.077 & 0.680 & Pass\\
2011 EU debt crisis & 0.020 & 0.081 & Fail: reverse also below 0.10\\
2015 China episode & 0.104 & 0.194 & Fail: forward above 0.10\\
2018 volatility shock & 0.055 & 0.559 & Pass\\
2020 COVID episode & 0.031 & 0.114 & Pass\\
2022 rate hikes & 0.711 & 0.049 & Fail\\\bottomrule
\end{tabular}
\caption{Original Table 5 values. Three of six rows satisfy its stated two-direction rule. Four satisfy only the forward $p<0.10$ criterion. None of these counts establishes independent out-of-sample validation.}\label{tab:events}
\end{table}
The 2015 row is not a significant success, and a relative ordering of two $p$-values does not determine an effect direction. These episodes also belong to the sample used for regime estimation and pair selection. They are retrospective subgroup checks. A binomial calculation treating selected episodes as independent Bernoulli validation trials has no justified null success probability here and should be removed.

\subsection{Alarms and economic interpretation}
The historical early-warning table measures time from a retrospectively identified regime assignment to a later selected volatility peak. It does not measure warning time before an unforeseen event onset. Requiring three consecutive crisis days also delays confirmation until the third day; a deployable alarm cannot be timestamped at the first day of that run without adjustment. Calendar-day and trading-day intervals must be labeled separately.

The original directional strategy reports annual return $-6.1\%$, Sharpe $-0.75$, and maximum drawdown $-97.5\%$. These values are unverified and its regime probabilities are described ambiguously as ``filtered (smoothed).'' They motivate a cautious interpretation but cannot isolate the economic value of the properly specified forecast model. A sign-of-nine-day-HML rule is not equivalent to forecasting from estimated distributed-lag coefficients.

The claim that a nine-day lag is stable across maximum lag limits including five is internally inconsistent as written. It may mean that significance survives a five-lag model, but the selected order cannot be nine when the cap is five. This statement requires reconstruction from the original model-selection outputs.

\section{Mechanism cost controls}
\begin{table}[ht]
\centering\small
\input{results/mechanism-full/mechanism-costs-table.tex}
\caption{Known-conditional risk costs with known-parameter gates. Parentheses are Monte Carlo standard errors over 100 independent seeds. OLS is minimum-norm least squares for predictions; redundant state blocks do not identify unique coefficients. The selected ridge comparison includes tuning and mean fallback.}\label{tab:mechanism-costs}
\end{table}
\begin{table}[ht]
\centering\footnotesize
\input{results/mechanism-full/mechanism-gate-costs-table.tex}
\caption{Fitted-gate minus known-gate ridge risk, with paired Monte Carlo standard errors. Costs include HMM approximation, numerical optimization and downstream retuning. Exact source-state observation makes this a favorable constructed gate-recovery setting; near-zero unrestricted costs are not evidence of general HMM consistency.}\label{tab:mechanism-gate-costs}
\end{table}

\subsection{Application-budget and horizon calibration}\label{app:matched-calibration}
The matched extension uses 12 scenarios, each with 2,000 independent datasets, $M=2400$ correlated comparisons, $n=416$, $\alpha=0.05$ and $\rho=1$. It matches the family budget and US temporal horizon only: all weights are one, outcomes follow the earlier two-point generator, and neither financial effect sizes nor heterogeneous regime-weight concentration are matched. Shared innovations induce comparison dependence. All-time family coverage is checked against exactly known cumulative conditional risks. All scenarios have zero observed coverage failures; a zero count in 2,000 has a pointwise 95\% Wilson upper endpoint of 0.191705\%. This is finite Monte Carlo evidence, not proof of zero error or exact nominal calibration.

\begin{table}[htbp]
\centering\small
\input{results/matched-inference-calibration/power-table.tex}
\caption{Matched-budget calibration. Crossing is positive-boundary crossing in any comparison or time: type-I error in null rows, descriptive detection in alternatives. Intervals are pointwise Wilson Monte Carlo intervals, not simultaneous or financial confidence intervals. Width is the terminal full normal-mixture width before support intersection; true gain is the simulated conditional target summary. Strong/weak are synthetic generator labels.}\label{tab:matched-calibration}
\end{table}
The independent and adaptive-variance strong-alternative counts are 1,356/2,000 and 928/2,000; persistent and abrupt-change strong alternatives have zero detections. Weak alternatives have zero detections throughout. These unfavorable cases are retained. Matching two design dimensions does not certify sensitivity for the empirical pipeline.

\subsection{Learned-pipeline conditional-risk diagnostic}\label{app:learned-inference}
This post-review diagnostic reuses all 400 original mechanism ledgers (100 seeds in each of four scenarios), after their MSE findings were inspected. The new capped-inference protocol was frozen before computing these statistics. No new seed or HMM fit is added. Each cap is $\max\{9\widehat{\operatorname{Var}}(Y_{0:399}),10^{-12}\}$. The full-past conditional outcome law is a two-component unit-variance Gaussian mixture; its means and weights integrate the next state while retaining the observed previous lag signal. Analytic capped moments were checked against numerical quadrature before execution.

Eleven paired comparisons (including exact-oracle, known-gate ridge/OLS, pooled and fitted-HMM procedures) are each evaluated with unit weights and the two past-only exact restricted state probabilities: 33 actual endpoints over 600 prefixes. Primary $M=2400$ assigns the financial-sized budget to these 33 endpoints; it is not 2,400 distinct simulated procedures. Secondary $M=33$ covers the natural family. These two displays are separate diagnostics, not a jointly corrected new family. Coverage failure means any interval misses its known cumulative conditional target over the 33 endpoints and all prefixes.

\begin{table}[htbp]
\centering\small
\input{results/learned-pipeline-inference/learned-inference-table.tex}
\caption{Learned-pipeline diagnostic. Failures count any-time family coverage misses in 100 seeds. Width averages support-intersected terminal widths over endpoints and seeds. Exact, Gaussian and Student columns count positive crossings of the unweighted information-exclusion comparison, not all endpoints. Zero failures in 100 has a pointwise 95\% Wilson upper endpoint of about 3.70\%.}\label{tab:learned-inference}
\end{table}
All unweighted exact-oracle, Gaussian and Student information-exclusion terminal targets are positive in every seed of the lag-only, gate-only and mixed scenarios, yet none is resolved positively under either allocation. For the exact oracle, their mean capped targets are 0.0421021, 0.00596873 and 0.0301856, respectively. Subsetting by positive realized conditional targets gives descriptive resolution fractions, not unconditional power estimates. Conditional-mean oracles minimize squared loss, not necessarily capped loss. In the null generator, fitted procedures need not have zero conditional capped differences; type-I labels apply only to the identically equal exact-oracle forecasts. Endpoint and weight dependence, shared seeds across scenarios and prior MSE inspection preclude treating these rows as independent confirmatory studies.

\section{Noisy-source stress-study details}\label{app:noisy-source}
The new seeds are 6000--6099, paired across all four corruption levels. A separate channel random stream leaves retained series and nuisance $L$ unchanged. Reused uniforms couple the noise levels without making their realized information sets nested. Population binary-channel garbling implies nonincreasing oracle information as corruption increases to $0.5$; finite coupled averages and fitted gains need not be monotone. The restricted filter remains the retained-observation filter, while the unrestricted filter multiplies retained emission likelihoods by $1-r$ or $r$ according to agreement with the proxy. Tests check small-state enumeration, predictive prefix invariance, endpoint identities and restricted-fit invariance. Pilot runs are stored separately and excluded.

\begin{table}[htbp]
\centering\small
\input{results/noisy-source-full/noisy-source-gains-table.tex}
\caption{Noisy-source known-conditional squared-loss gains. Each row averages 100 independent seeds; parentheses are Monte Carlo standard errors. Seeds are paired across noise levels. Exact shared-gate gains are zero by construction. These are constructed block-source comparisons, not financial confidence intervals.}\label{tab:noisy-source-gains}
\end{table}
\begin{table}[htbp]
\centering\small
\input{results/noisy-source-full/noisy-source-paired-changes-table.tex}
\caption{Paired changes from the uncorrupted-source setting. Pairing preserves retained observations; uncertainty describes independent seed averages rather than simultaneous inference across rows.}\label{tab:noisy-source-changes}
\end{table}
\begin{table}[htbp]
\centering\small
\input{results/noisy-source-full/noisy-source-gate-costs-table.tex}
\caption{Fitted-gate minus oracle-gate ridge risk with paired Monte Carlo uncertainty. Binary-proxy emission misspecification, estimation, optimization and downstream selection are combined; the contrasts do not isolate pure gate-estimation error.}\label{tab:noisy-source-costs}
\end{table}
All 400 ledgers and aggregate contrasts were verified, including exact restricted forecast/penalty invariance across paired levels. The nonconvergence count is zero, which certifies the implemented stopping criterion only. The source remains a two-coordinate block with one irrelevant coordinate. Negative fitted gains at the independent-channel endpoint are finite-procedure failures, not violations of nonnegative population oracle information.

\section{Estimated-gate implementation details}\label{app:emissions}
Let $z_t\in\{1,\ldots,K\}$ have transition probabilities $A_{jk}$, and let observations $\mathbf x_t$ follow a multivariate Student-$t$ emission:
\begin{equation}
 p(\mathbf x\mid z=k)=
 \frac{\Gamma((\nu_k+d)/2)}{\Gamma(\nu_k/2)(\nu_k\pi)^{d/2}|\Sigma_k|^{1/2}}
 \left[1+\frac{(\mathbf x-\mu_k)^\top\Sigma_k^{-1}(\mathbf x-\mu_k)}{\nu_k}\right]^{-(\nu_k+d)/2}.
\end{equation}
Here $\Sigma_k$ is a scale matrix; when $\nu_k>2$, covariance equals $\nu_k\Sigma_k/(\nu_k-2)$. Treating the two matrices as identical misstates volatility comparisons. The executed model fixes $\nu=5$ and diagonal scales; initialization, scale floors and convergence settings are documented in the computational appendix.

The HMM estimator is a new reference implementation of established EM machinery, not a new statistical method or a recovery of the original code. For diagonal Student-$t$ scale $D_k$, dimension $d$, and fixed $\nu=5$, define $\delta_{tk}=(x_t-\mu_k)^\top D_k^{-1}(x_t-\mu_k)$. The E-step computes smoothed occupancies $\gamma_{tk}$ and adjacent-state masses $\xi_{tjk}$ within the fit sample. Its latent scale expectation is $u_{tk}=(\nu+d)/(\nu+\delta_{tk})$. The M-step updates
\begin{align}
 \mu_k^{\rm new}&=\frac{\sum_t\gamma_{tk}u_{tk}x_t}{\sum_t\gamma_{tk}u_{tk}},\\
 D_{k,aa}^{\rm new}&=\frac{\sum_t\gamma_{tk}u_{tk}(x_{ta}-\mu_{ka}^{\rm new})^2}{\sum_t\gamma_{tk}},\\
 A_{jk}^{\rm new}&=\frac{\sum_{t=1}^{T-1}\xi_{tjk}}{\sum_m\sum_{t=1}^{T-1}\xi_{tjm}}.
\end{align}
The Gaussian update uses $u_{tk}=1$. The initial-state distribution is updated from $\gamma_{1k}$ rather than constrained to the stationary distribution. A diagonal multivariate Student-$t$ scale does not make its coordinates independent; they retain a shared radial mixing variable.

Features are standardized with fit-sample means and standard deviations (floor $10^{-8}$). Diagonal scales are bounded below by $10^{-4}$ in standardized units, and transition/initial masses are clipped at $10^{-12}$ before normalization. The first initialization partitions observations by squared standardized norm; the second uses seeded random observations as means with perturbed common scales. Both start from $0.9I+0.1\mathbf1\mathbf1^\top/K$. We select the larger final training log likelihood and stop when successive likelihood changes are at most $10^{-5}(1+|\ell_{\rm previous}|)$ or the iteration cap is reached. Fit diagnostics retain convergence and likelihood-decrease counts. States are ordered by training covariance trace in original units; no evaluation-event labels enter this ordering.

The smoothed probabilities used to estimate parameters are never used as evaluation gates. The fixed-parameter forward filter supplies predictive weights. Tests compare emission densities with SciPy, forward--backward results with exhaustive enumeration of a small state space, optional compiled recursions with Python, and current/future perturbations with unchanged previously issued forecasts. Complete source-exclusion tests also perturb the source over the entire dataset and require restricted predictions and selected penalties to remain identical. These establish specific implementation properties, not global statistical validity.

\section{An elementary binary-channel concentration mechanism}\label{app:binary-channel}
The following likelihood-ratio argument describes a specified Gaussian-emission parameter sequence. It is not a theorem about all maximum-likelihood solutions or EM trajectories. Modeling discrete and continuous coordinates with their respective emission distributions is established HMM practice \cite{visser2010}.

\begin{proposition}[Concentration on an observed proxy]
Fix observations $(z_t,x_t)_{t=1}^n$, with $x_t\in\{-1,1\}$. Let model states $K_t\in\{-1,1\}$ have strictly positive initial probabilities $\pi_k$ and transitions $A_{kl}$, and emissions
\[
 f_{k,\varepsilon}(z_t,x_t)=g_k(z_t)\phi(x_t;k,\varepsilon^2),
\]
where every observed $g_k(z_t)$ is finite and positive. Set
\[
 R_\pi=\frac{\max_k\pi_k}{\min_k\pi_k},\qquad
 R_A=\frac{\max_{kl}A_{kl}}{\min_{kl}A_{kl}},\qquad
 G=\max\left\{1,\max_t\frac{g_{-x_t}(z_t)}{g_{x_t}(z_t)}\right\}.
\]
Writing $a_\varepsilon=R_A^2G\exp(-2/\varepsilon^2)$ and
$B_\varepsilon=R_\pi\{(1+a_\varepsilon)^n-1\}$, the model posterior satisfies
\[
 P_\varepsilon(K_{1:n}\ne x_{1:n}\mid z_{1:n},x_{1:n})
 \le \frac{B_\varepsilon}{1+B_\varepsilon}.
\]
For fixed $n$ and uniformly bounded displayed ratios, this probability tends to zero as $\varepsilon\downarrow0$.
\end{proposition}
\begin{proof}
Let $W(k)$ denote the unnormalized posterior weight of a state path. A path differing from $x$ at $h$ positions incurs Gaussian density ratio $\exp(-2h/\varepsilon^2)$. Its other emission ratios contribute at most $G^h$. At most $2h$ transition edges touch a changed position, so transition and initial ratios are bounded by $R_A^{2h}R_\pi$. Thus $W(k)/W(x)\le R_\pi a_\varepsilon^h$. There are ${n\choose h}$ binary paths with $h$ mismatches. Summing gives total alternative/reference weight ratio at most $B_\varepsilon$; normalization yields the result.
\end{proof}

\paragraph{Forecast timing.}
This smoothing statement does not license using current or future observations in a forecast. If $\alpha_t$ is the filtered distribution after assimilating observation $t$, its odds of the mismatched state are bounded, for $t\ge2$, by
\[
 \frac{\alpha_t(-x_t)}{\alpha_t(x_t)}
 \le d_\varepsilon:=R_AG\exp(-2/\varepsilon^2).
\]
Indeed, the predictive probabilities before observation $t$ lie between the smallest and largest transition entries. The next predictive gate therefore obeys
\[
 \operatorname{TV}(\alpha_tA,A_{x_t,\cdot})
 =\alpha_t(-x_t)\operatorname{TV}(A_{-x_t,\cdot},A_{x_t,\cdot})
 \le\frac{d_\varepsilon}{1+d_\varepsilon}.
\]
At $t=1$, replace $R_A$ by $R_\pi$. With fixed transitions, the predictive gate approaches the row indexed by the previous observed proxy; it need not approach a point mass or the true latent state. For varying transitions the bound concerns the corresponding $A_\varepsilon$ row, whose convergence is an additional assumption. Chronological validity also requires past-only fitting, standardization and model selection.

\paragraph{What the result can explain.}
If the observed channel is a noisy proxy of another state $S_t$, concentration of model state $K_t$ on $X_t$ does not imply recovery of $S_t$. A correctly specified Bernoulli channel with fixed corruption $r\in(0,1/2)$ has finite likelihood ratio $(1-r)/r$, whereas the displayed Gaussian sequence has ratio $\exp(2/\varepsilon^2)$. This identifies one possible density incentive; it does not establish a forecast-risk sign. The experiment estimates channel probabilities with a positive floor and Gaussian means and variances with a fixed floor. Its fitted outputs therefore need direct empirical inspection. The proposition pins means to the two observed support values and does not cover simultaneous degeneration of other channels or exponentially vanishing transition probabilities. Under standardization, the exponent becomes the squared support separation divided by twice the variance in the same units. Gaussian density and Bernoulli probability-mass likelihood values are never compared directly.

\end{document}

%% file: preamble.tex
\documentclass[11pt,a4paper]{article}
\usepackage[margin=25mm]{geometry}
\usepackage[T1]{fontenc}
\usepackage{lmodern,microtype,amsmath,amssymb,amsthm,booktabs,tabularx,array,graphicx,xcolor,enumitem,fancyhdr}
\usepackage[section]{placeins}
\usepackage{float}
\usepackage[hidelinks]{hyperref}
\definecolor{ink}{HTML}{17324D}
\definecolor{muted}{HTML}{526475}
\definecolor{wash}{HTML}{EDF3F7}
\newcommand{\E}{\mathbb{E}}
\newcommand{\Prb}{\mathbb{P}}
\newcommand{\R}{\mathbb{R}}

\newtheorem{proposition}{Proposition}

\theoremstyle{definition}\newtheorem{example}{Example}
\setlist{nosep,leftmargin=1.5em}
\newcommand{\sourceurl}[1]{\url{#1}}

%% file: results/panel-comparison/family-table.tex
\begin{tabular}{llrrrrr}\toprule
Region & Period & $N$ & Positive & Negative & Margin & Identical\\\midrule
Asia\_Pacific\_ex\_Japan & historical & 600 & 0 & 0 & 362 & 0\\
Asia\_Pacific\_ex\_Japan & temporal & 600 & 0 & 0 & 102 & 0\\
Europe & historical & 600 & 0 & 0 & 402 & 16\\
Europe & temporal & 600 & 0 & 0 & 165 & 16\\
Japan & historical & 600 & 0 & 0 & 426 & 76\\
Japan & temporal & 600 & 0 & 0 & 197 & 76\\
US & historical & 600 & 0 & 0 & 460 & 104\\
US & temporal & 600 & 0 & 0 & 312 & 104\\
\bottomrule
\end{tabular}

%% file: results/mechanism-full/mechanism-gains-table.tex
\begin{tabular}{llrr}\toprule
Scenario & Procedure & Information gain & Shared-gate gain\\\midrule
Null & Exact oracle & 0.0000 (0.0000) & 0.0000 (0.0000)\\
Null & Pooled ridge & -0.0025 (0.0005) & ---\\
Null & Oracle-gate ridge & -0.0024 (0.0007) & -0.0022 (0.0004)\\
Null & Oracle-gate OLS & -0.0030 (0.0005) & -0.0029 (0.0003)\\
Null & Gaussian gate & -0.0026 (0.0009) & -0.0022 (0.0004)\\
Null & Student-$t$ gate & -0.0025 (0.0009) & -0.0022 (0.0004)\\
\addlinespace
Lag only & Exact oracle & 0.6457 (0.0036) & 0.6457 (0.0036)\\
Lag only & Pooled ridge & 0.6498 (0.0039) & ---\\
Lag only & Oracle-gate ridge & 0.6533 (0.0042) & 0.6523 (0.0038)\\
Lag only & Oracle-gate OLS & 0.6482 (0.0038) & 0.6480 (0.0038)\\
Lag only & Gaussian gate & 0.6539 (0.0043) & 0.6525 (0.0038)\\
Lag only & Student-$t$ gate & 0.6538 (0.0043) & 0.6525 (0.0038)\\
\addlinespace
Gate only & Exact oracle & 0.1301 (0.0027) & 0.0000 (0.0000)\\
Gate only & Pooled ridge & 0.3061 (0.0021) & ---\\
Gate only & Oracle-gate ridge & 0.1284 (0.0029) & -0.0036 (0.0004)\\
Gate only & Oracle-gate OLS & 0.1271 (0.0028) & -0.0039 (0.0004)\\
Gate only & Gaussian gate & 0.1345 (0.0032) & -0.0036 (0.0004)\\
Gate only & Student-$t$ gate & 0.1374 (0.0032) & -0.0036 (0.0004)\\
\addlinespace
Mixed & Exact oracle & 0.8355 (0.0059) & 0.6457 (0.0036)\\
Mixed & Pooled ridge & 1.0432 (0.0061) & ---\\
Mixed & Oracle-gate ridge & 0.8467 (0.0062) & 0.6560 (0.0052)\\
Mixed & Oracle-gate OLS & 0.8382 (0.0058) & 0.6476 (0.0039)\\
Mixed & Gaussian gate & 0.8575 (0.0064) & 0.6559 (0.0052)\\
Mixed & Student-$t$ gate & 0.8606 (0.0065) & 0.6559 (0.0052)\\
\addlinespace
\bottomrule\end{tabular}

%% file: results/emission-intervention-full/emission-intervention-table.tex
\begin{tabular}{lrrrl}\toprule
$r$ & $\Delta$ U risk & $\Delta$ Brier & $\Delta$ log score & Floor fits \\\midrule
0.0 & \shortstack[r]{0.000208\\{\scriptsize [-8.15e-06, 0.000425]}} & \shortstack[r]{5.65e-05\\{\scriptsize [-3.76e-06, 0.000117]}} & \shortstack[r]{0.000183\\{\scriptsize [-2.15e-05, 0.000388]}} & 400/400\\
0.1 & \shortstack[r]{-0.123\\{\scriptsize [-0.128, -0.117]}} & \shortstack[r]{-0.0637\\{\scriptsize [-0.0656, -0.0618]}} & \shortstack[r]{-0.183\\{\scriptsize [-0.189, -0.178]}} & 394/400\\
0.3 & \shortstack[r]{0.000204\\{\scriptsize [-0.00163, 0.00204]}} & \shortstack[r]{-0.000192\\{\scriptsize [-0.000292, -9.22e-05]}} & \shortstack[r]{-0.00048\\{\scriptsize [-0.000753, -0.000207]}} & 2/400\\
0.5 & \shortstack[r]{3.23e-06\\{\scriptsize [-4.47e-06, 1.09e-05]}} & \shortstack[r]{3.03e-06\\{\scriptsize [-5.95e-06, 1.2e-05]}} & \shortstack[r]{1.18e-05\\{\scriptsize [-1.7e-05, 4.06e-05]}} & 0/400\\
\bottomrule\end{tabular}

%% file: results/variance-adaptive-learned/paired-learned-table.tex
\begin{tabular}{llrrrr}\toprule
Scenario & $M$ & Fail N/E & Pos N/E & Neg N/E & Width N/E\\\midrule
null & 2400 & 0/0 & 0/0 & 0/0 & 0.02464/0.01535\\
null & 33 & 0/0 & 0/0 & 0/0 & 0.01979/0.01139\\
lag\_only & 2400 & 0/0 & 0/0 & 0/0 & 0.18786/0.13102\\
lag\_only & 33 & 0/0 & 0/50 & 0/0 & 0.15290/0.10399\\
gate\_only & 2400 & 0/0 & 0/0 & 0/0 & 0.05202/0.05495\\
gate\_only & 33 & 0/0 & 0/0 & 0/0 & 0.04197/0.04015\\
mixed & 2400 & 0/0 & 0/3 & 0/0 & 0.15955/0.11104\\
mixed & 33 & 0/0 & 0/86 & 0/0 & 0.12941/0.08408\\
\bottomrule
\end{tabular}

%% file: results/empirical-comparison/rmse-table.tex
\begin{tabular}{lrr}\toprule
Model & Two lags & Nine lags\\\midrule
Zero return & 63.948 & 63.948\\
Historical mean & 63.954 & 63.954\\
Own lags & 63.977 & 64.035\\
Pooled, exclude HML & 64.177 & 64.359\\
Pooled, include HML & 64.248 & 64.438\\
Volatility gate, include HML & 64.282 & 64.639\\
Gaussian HMM, exclude HML & 64.319 & 64.604\\
Gaussian HMM, include HML & 64.403 & 64.660\\
Student-$t$ HMM, exclude HML & 64.257 & 64.540\\
Student-$t$ HMM, include HML & 64.322 & 64.622\\
\bottomrule
\end{tabular}

%% file: results/empirical-comparison/regime-table.tex
\begin{tabular}{lrrrr}\toprule
State rank & Weight mass & Pooled & Gaussian & Student-$t$\\\midrule
Low & 1147.4 & -1.49 & 1.65 & -2.87\\
Middle & 1888.6 & -6.88 & -9.36 & -6.68\\
High & 737.9 & -26.69 & -34.11 & -21.46\\
\bottomrule
\end{tabular}

%% file: results/empirical-comparison/audit-sensitivity-table.tex
\begin{tabular}{llrr}\toprule
Gate & Coding & Exclusion gain & Shared-gate gain\\\midrule
Gaussian & Low reference & -10.129 & -10.044\\
Gaussian & Middle reference & -0.224 & -0.291\\
Gaussian & High reference & -0.158 & -0.195\\
Gaussian & Symmetric & -0.218 & -0.229\\
Student & Low reference & -7.015 & -7.019\\
Student & Middle reference & -0.161 & -0.299\\
Student & High reference & -1.499 & -4.553\\
Student & Symmetric & -3.635 & -3.747\\
\bottomrule
\end{tabular}

%% file: results/empirical-comparison/simulation-table.tex
\begin{tabular}{lrrr}\toprule
Process & Pooled & Gaussian HMM & Student-$t$ HMM\\\midrule
Gaussian null & -0.0041 (0.0014) & -0.0037 (0.0017) & -0.0021 (0.0011)\\
Linear signal & 0.6290 (0.0202) & 0.6219 (0.0194) & 0.6201 (0.0205)\\
Heavy-tailed null & -0.0039 (0.0012) & -0.0033 (0.0015) & -0.0016 (0.0017)\\
Switching signal & -0.0162 (0.0082) & 0.5601 (0.0271) & 0.5462 (0.0259)\\
\bottomrule
\end{tabular}

%% file: results/mechanism-full/mechanism-costs-table.tex
\begin{tabular}{llrr}\toprule
Scenario & Path & OLS minus exact & Ridge minus OLS\\\midrule
Null & R & 0.0088 (0.0005) & 0.0033 (0.0009)\\
Null & U & 0.0118 (0.0007) & 0.0027 (0.0008)\\
Null & shared\_R & 0.0089 (0.0006) & 0.0034 (0.0008)\\
Lag only & R & 0.0141 (0.0010) & 0.0064 (0.0017)\\
Lag only & U & 0.0116 (0.0006) & 0.0013 (0.0003)\\
Lag only & shared\_R & 0.0140 (0.0009) & 0.0056 (0.0012)\\
Gate only & R & 0.0113 (0.0008) & 0.0054 (0.0021)\\
Gate only & U & 0.0143 (0.0008) & 0.0041 (0.0021)\\
Gate only & shared\_R & 0.0104 (0.0006) & 0.0044 (0.0021)\\
Mixed & R & 0.0169 (0.0012) & 0.0101 (0.0029)\\
Mixed & U & 0.0142 (0.0007) & 0.0016 (0.0004)\\
Mixed & shared\_R & 0.0161 (0.0011) & 0.0100 (0.0033)\\
\bottomrule\end{tabular}

%% file: results/mechanism-full/mechanism-gate-costs-table.tex
\begin{tabular}{llrr}\toprule
Scenario & Path & Gaussian minus oracle gate & Student-$t$ minus oracle gate\\\midrule
Null & R & -0.0002 (0.0006) & -0.0001 (0.0007)\\
Null & U & 0.0000 (0.0000) & 0.0000 (0.0000)\\
Null & shared\_R & 0.0000 (0.0000) & 0.0000 (0.0000)\\
Lag only & R & 0.0006 (0.0007) & 0.0005 (0.0008)\\
Lag only & U & -0.0000 (0.0000) & -0.0000 (0.0000)\\
Lag only & shared\_R & 0.0002 (0.0002) & 0.0002 (0.0002)\\
Gate only & R & 0.0061 (0.0011) & 0.0091 (0.0013)\\
Gate only & U & 0.0000 (0.0000) & 0.0000 (0.0000)\\
Gate only & shared\_R & 0.0000 (0.0000) & 0.0000 (0.0000)\\
Mixed & R & 0.0108 (0.0016) & 0.0140 (0.0019)\\
Mixed & U & 0.0000 (0.0000) & 0.0000 (0.0000)\\
Mixed & shared\_R & -0.0000 (0.0000) & -0.0000 (0.0000)\\
\bottomrule\end{tabular}

%% file: results/matched-inference-calibration/power-table.tex
\begin{tabular}{llrrrr}\toprule
Environment & Signal & Crossing (\%) & 95\% MC interval & Width & True gain\\\midrule
IID & Null & 0.00 & [0.00, 0.19] & 0.2372 & 0.0000\\
IID & Weak & 0.00 & [0.00, 0.19] & 0.2372 & 0.0183\\
IID & Strong & 67.80 & [65.72, 69.81] & 0.2372 & 0.1222\\
\addlinespace
Persistent & Null & 0.00 & [0.00, 0.19] & 0.2551 & 0.0000\\
Persistent & Weak & 0.00 & [0.00, 0.19] & 0.2548 & 0.0094\\
Persistent & Strong & 0.00 & [0.00, 0.19] & 0.2552 & 0.0625\\
\addlinespace
Abrupt & Null & 0.00 & [0.00, 0.19] & 0.2603 & 0.0000\\
Abrupt & Weak & 0.00 & [0.00, 0.19] & 0.2603 & 0.0096\\
Abrupt & Strong & 0.00 & [0.00, 0.19] & 0.2603 & 0.0640\\
\addlinespace
Adaptive variance & Null & 0.00 & [0.00, 0.19] & 0.2399 & 0.0000\\
Adaptive variance & Weak & 0.00 & [0.00, 0.19] & 0.2410 & 0.0165\\
Adaptive variance & Strong & 46.40 & [44.22, 48.59] & 0.2392 & 0.1086\\
\addlinespace
\bottomrule
\end{tabular}

%% file: results/learned-pipeline-inference/learned-inference-table.tex
\begin{tabular}{llrrrrr}\toprule
Scenario & $M$ & Failures & Width & Exact & Gaussian & Student\\\midrule
null & 2400 & 0/100 & 0.0246 & 0/100 & 0/100 & 0/100\\
null & 33 & 0/100 & 0.0198 & 0/100 & 0/100 & 0/100\\
lag\_only & 2400 & 0/100 & 0.1879 & 0/100 & 0/100 & 0/100\\
lag\_only & 33 & 0/100 & 0.1529 & 0/100 & 0/100 & 0/100\\
gate\_only & 2400 & 0/100 & 0.0520 & 0/100 & 0/100 & 0/100\\
gate\_only & 33 & 0/100 & 0.0420 & 0/100 & 0/100 & 0/100\\
mixed & 2400 & 0/100 & 0.1595 & 0/100 & 0/100 & 0/100\\
mixed & 33 & 0/100 & 0.1294 & 0/100 & 0/100 & 0/100\\
\bottomrule
\end{tabular}

%% file: results/noisy-source-full/noisy-source-gains-table.tex
\begin{tabular}{llrr}\toprule
Flip probability & Procedure & Information gain & Shared-gate gain\\\midrule
0.0 & Exact oracle & 0.1336 (0.0026) & 0.0000 (0.0000)\\
0.0 & Pooled ridge & 0.3070 (0.0028) & ---\\
0.0 & Oracle-gate ridge & 0.1317 (0.0027) & -0.0039 (0.0011)\\
0.0 & Gaussian gate & 0.1386 (0.0030) & -0.0039 (0.0011)\\
0.0 & Student-$t$ gate & 0.1419 (0.0029) & -0.0039 (0.0011)\\
\addlinespace
0.1 & Exact oracle & 0.0708 (0.0013) & 0.0000 (0.0000)\\
0.1 & Pooled ridge & 0.1229 (0.0017) & ---\\
0.1 & Oracle-gate ridge & 0.0690 (0.0018) & -0.0034 (0.0008)\\
0.1 & Gaussian gate & -0.0494 (0.0024) & -0.0048 (0.0011)\\
0.1 & Student-$t$ gate & -0.0486 (0.0023) & -0.0047 (0.0011)\\
\addlinespace
0.3 & Exact oracle & 0.0174 (0.0004) & 0.0000 (0.0000)\\
0.3 & Pooled ridge & 0.0183 (0.0007) & ---\\
0.3 & Oracle-gate ridge & 0.0131 (0.0017) & -0.0045 (0.0011)\\
0.3 & Gaussian gate & -0.0216 (0.0041) & -0.0048 (0.0012)\\
0.3 & Student-$t$ gate & -0.0327 (0.0044) & -0.0034 (0.0009)\\
\addlinespace
0.5 & Exact oracle & 0.0000 (0.0000) & 0.0000 (0.0000)\\
0.5 & Pooled ridge & -0.0035 (0.0004) & ---\\
0.5 & Oracle-gate ridge & -0.0034 (0.0008) & -0.0034 (0.0008)\\
0.5 & Gaussian gate & -0.0336 (0.0041) & -0.0040 (0.0011)\\
0.5 & Student-$t$ gate & -0.0446 (0.0038) & -0.0030 (0.0006)\\
\addlinespace
\bottomrule\end{tabular}

%% file: results/noisy-source-full/noisy-source-paired-changes-table.tex
\begin{tabular}{lrrrr}\toprule
Flip probability & Oracle & Oracle-gate ridge & Gaussian & Student-$t$\\\midrule
0.0 & 0.0000 (0.0000) & 0.0000 (0.0000) & 0.0000 (0.0000) & 0.0000 (0.0000)\\
0.1 & -0.0628 (0.0017) & -0.0627 (0.0018) & -0.1880 (0.0032) & -0.1906 (0.0029)\\
0.3 & -0.1162 (0.0025) & -0.1186 (0.0030) & -0.1602 (0.0051) & -0.1747 (0.0053)\\
0.5 & -0.1336 (0.0026) & -0.1351 (0.0027) & -0.1722 (0.0050) & -0.1865 (0.0051)\\
\bottomrule\end{tabular}

%% file: results/noisy-source-full/noisy-source-gate-costs-table.tex
\begin{tabular}{llrr}\toprule
Flip probability & Path & Gaussian minus oracle gate & Student-$t$ minus oracle gate\\\midrule
0.0 & R & 0.0069 (0.0011) & 0.0102 (0.0013)\\
0.0 & U & 0.0000 (0.0000) & 0.0000 (0.0000)\\
0.0 & shared\_R & -0.0000 (0.0000) & -0.0000 (0.0000)\\
0.1 & R & 0.0066 (0.0011) & 0.0100 (0.0012)\\
0.1 & U & 0.1250 (0.0021) & 0.1276 (0.0019)\\
0.1 & shared\_R & 0.1236 (0.0022) & 0.1262 (0.0019)\\
0.3 & R & 0.0061 (0.0010) & 0.0097 (0.0012)\\
0.3 & U & 0.0408 (0.0044) & 0.0556 (0.0049)\\
0.3 & shared\_R & 0.0406 (0.0043) & 0.0567 (0.0048)\\
0.5 & R & 0.0063 (0.0010) & 0.0096 (0.0012)\\
0.5 & U & 0.0364 (0.0042) & 0.0508 (0.0040)\\
0.5 & shared\_R & 0.0358 (0.0042) & 0.0513 (0.0042)\\
\bottomrule\end{tabular}

%% file: main.bbl
\begin{thebibliography}{99}
\small
\setlength{\itemsep}{3pt}
\bibitem{visser2010} I. Visser and M. Speekenbrink. depmixS4: An R Package for Hidden Markov Models. Journal of Statistical Software, 36(7):1--21, 2010. \url{https://doi.org/10.18637/jss.v036.i07}.
\bibitem{howard2021} S. R. Howard, A. Ramdas, J. McAuliffe, and J. Sekhon. Time-uniform, nonparametric, nonasymptotic confidence sequences. Annals of Statistics, 49(2):1055--1080, 2021. \url{https://doi.org/10.1214/20-AOS1991}.
\bibitem{choe2024} Y. J. Choe and A. Ramdas. Comparing Sequential Forecasters. Operations Research, 72(4):1368--1387, 2024. \url{https://doi.org/10.1287/opre.2021.0792}.
\bibitem{arnold2026} S. Arnold, G. Gavrilopoulos, B. Schulz, and J. Ziegel. Sequential model confidence sets. arXiv:2404.18678v4, 2026. \url{https://arxiv.org/abs/2404.18678}.

\bibitem{covert} I. Covert, S. Lundberg, and S.-I. Lee. Explaining by Removing: A Unified Framework for Model Explanation. Journal of Machine Learning Research, 22(209):1--90, 2021. \url{https://www.jmlr.org/papers/v22/20-1316.html}.
\bibitem{tank} A. Tank, I. Covert, N. Foti, A. Shojaie, and E. Fox. Neural Granger Causality. IEEE Transactions on Pattern Analysis and Machine Intelligence, accepted version, 2021. \url{https://doi.org/10.1109/TPAMI.2021.3065601}.
\bibitem{frenchdeveloped} K. R. French. Description of Fama/French 5 Factors for Developed Markets. Accessed 26 September 2026. \url{https://mba.tuck.dartmouth.edu/pages/faculty/ken.french/Data_Library/f-f_5developed.html}.
\bibitem{frenchmom} K. R. French. Description of Daily Momentum Factors for Developed Markets. Accessed 26 September 2026. \url{https://mba.tuck.dartmouth.edu/pages/faculty/ken.french/Data_Library/f-f_developed_daily_mom.html}.
\bibitem{clarkwest} T. E. Clark and K. D. West. Approximately normal tests for equal predictive accuracy in nested models. Journal of Econometrics, 138(1):291--311, 2007. \url{https://doi.org/10.1016/j.jeconom.2006.05.023}.
\bibitem{giacomini} R. Giacomini and H. White. Tests of Conditional Predictive Ability. Econometrica, 74(6):1545--1578, 2006. \url{https://doi.org/10.1111/j.1468-0262.2006.00718.x}.
\bibitem{politis} D. N. Politis and J. P. Romano. The Stationary Bootstrap. Journal of the American Statistical Association, 89(428):1303--1313, 1994. \url{https://doi.org/10.1080/01621459.1994.10476870}.
\bibitem{droumaguet} M. Droumaguet, A. Warne, and T. Wo\'zniak. Granger causality and regime inference in Bayesian Markov-Switching VARs. Journal of Applied Econometrics, 32(4):802--818, 2017. \url{https://doi.org/10.1002/jae.2531}. Accessible version: ECB Working Paper 1794, 2015. \url{https://www.ecb.europa.eu/pub/pdf/scpwps/ecbwp1794.en.pdf}.
\bibitem{original} C. Lee. Regime-Dependent Predictive Structure Between Equity Factors: Evidence from Granger Causality. arXiv:2601.10732v1, 2026. \url{https://arxiv.org/abs/2601.10732v1}.
\bibitem{granger} C. W. J. Granger. Investigating Causal Relations by Econometric Models and Cross-spectral Methods. Econometrica, 37(3):424--438, 1969. \url{https://doi.org/10.2307/1912791}.
\bibitem{bulla} J. Bulla. Hidden Markov models with t components. Increased persistence and other aspects. Quantitative Finance, 11(3):459--475, 2011. \url{https://doi.org/10.1080/14697681003685563}.
\bibitem{french} K. R. French. Data Library. Primary data source, accessed 26 September 2026. \url{https://mba.tuck.dartmouth.edu/pages/faculty/ken.french/data_library.html}.
\end{thebibliography}
